\documentclass[notitlepage,11pt,superscriptaddress, onecolumn]{revtex4-1}

\usepackage{ntheorem}
\usepackage{enumerate}
\usepackage{amsmath}
\usepackage{amssymb}
\usepackage[dvipdfmx]{graphicx}
\usepackage{color}
\usepackage{euscript}
\usepackage{float}
\usepackage{hyperref}
\usepackage{hhline}
\usepackage{multirow}
\usepackage{relsize}

\theorembodyfont{\rmfamily}
\newtheorem{Theorem}{\textit{Theorem} }
\newtheorem{Definition}[Theorem]{\textit{Definition} }
\newtheorem{Corollary}[Theorem]{\textit{Corollary}}
\newtheorem{Lemma}[Theorem]{\textit{Lemma}}
\newtheorem*{Proof}{\textit{Proof}}
\newtheorem*{Conjecture}{\textit{Conjecture}}

\newcommand{\lv}{\left \vert}
\newcommand{\rv}{\right \vert}
\newcommand{\la}{\left \langle}
\newcommand{\ra}{\right \rangle}
\newcommand{\ket}[1]{\lv #1 \ra}
\newcommand{\bra}[1]{\la #1 \rv}
\newcommand{\braket}[2]{\la #1 \vert #2 \ra}
\newcommand{\ketbra}[2]{\lv #1 \ra \la #2 \rv}

\newcommand{\tr}{\mathrm{tr}}
\newcommand{\mc}[1]{\mathcal{#1}}
\newcommand{\mbf}[1]{\mathbf{#1}}
\newcommand{\blds}[1]{\boldsymbol{#1}}
\newcommand{\mbb}[1]{\mathbb{#1}}

\newcommand{\place}[2]{%
  \overset{\substack{#1\\\smile}}{#2}%
}

\begin{document}

\title{Efficient unitary designs with nearly time-independent Hamiltonian dynamics}

\author{Yoshifumi Nakata}
\affiliation{Photon Science Center, Graduate School of Engineering, The University of Tokyo, Bunkyo-ku, Tokyo 113-8656, Japan.}
\affiliation{ Departament de F\'{\i}sica: Grup d'Informaci\'{o} Qu\`{a}ntica, Universitat 
Aut\`{o}noma de Barcelona, ES-08193 Bellaterra (Barcelona), Spain}

\author{Christoph Hirche}
\affiliation{ Departament de F\'{\i}sica: Grup d'Informaci\'{o} Qu\`{a}ntica, Universitat 
Aut\`{o}noma de Barcelona, ES-08193 Bellaterra (Barcelona), Spain}

\author{Masato Koashi}
\affiliation{Photon Science Center, Graduate School of Engineering, The University of Tokyo, Bunkyo-ku, Tokyo 113-8656, Japan.}

\author{Andreas Winter}
\affiliation{ Departament de F\'{\i}sica: Grup d'Informaci\'{o} Qu\`{a}ntica, Universitat 
Aut\`{o}noma de Barcelona, ES-08193 Bellaterra (Barcelona), Spain}
\affiliation{ICREA: Instituci\'{o} Catalana de Recerca i Estudis Avan\c{c}ats, ES-08010 Barcelona, Spain.}

\begin{abstract}
We provide new constructions of unitary $t$-designs for general $t$ on one qudit and $N$ qubits, and propose a \emph{design Hamiltonian}, a random Hamiltonian of which dynamics always forms a unitary design after a threshold time, as a basic framework to investigate randomising time evolution in quantum many-body systems.
The new constructions are based on recently proposed schemes of repeating random unitaires diagonal in mutually unbiased bases. We first show that, if a pair of the bases satisfies a certain condition, the process on one qudit approximately forms a unitary $t$-design after $O(t)$ repetitions. We then construct quantum circuits on $N$ qubits that achieve unitary $t$-designs for $t = o(N^{1/2})$ using $O(t N^2)$ gates, improving the previous result using $O(t^{10}N^2)$ gates in terms of $t$.
Based on these results, we present a design Hamiltonian with periodically changing two-local spin-glass-type interactions, leading to fast and relatively natural realisations of unitary designs in complex many-body systems.
\end{abstract}

\maketitle

\section{Introduction}
Random quantum processes play important roles in quantum information processing, as one of the fundamental primitives in quantum Shannon theory~\cite{D2005,DW2004,GPW2005,ADHW2009,HaydenTutorial,DBWR2010,SDTR2013,HM14}
and as a useful resource to demonstrate quantum advantages in many protocols~\cite{EAZ2005,KLRetc2008,MGE2011,MGE2012,S2005,BH2013,KRT2014,KL15,KZD2016,OAGKAL2016}.
In recent years, random processes have been also revealed to be the key to understanding fundamental physics in complex quantum systems, leading to new developments in quantum thermodynamics~\cite{PSW2006,GLTZ2006,R2008} (see Ref.~\cite{GE2016} for a comprehensive review), black hole information science~\cite{HP2007,SS2008,S2011,LSHOH2013,S2014,HQRY2016} and strongly correlated many-body physics~\cite{SS2014,RD2015,SS2015}.
In quantum systems, random processes are often represented by random unitaries drawn uniformly at random according to the Haar measure, referred to as \emph{Haar random unitaries}.
However, when a system consists of a large number of particles, it is highly inefficient to implement Haar random unitaries,
implying that they rarely appear in natural systems composed of many particles especially when the interactions are local.
This fact has lead to the research area on finite-degree approximations of Haar random unitaries, so-called \emph{unitary designs}~\cite{DLT2002,DCEL2009,GAE2007}, and their efficient implementations~\cite{TGJ2007,BWV2008a,WBV2008,HL2009,DJ2011,HL2009TPE,BHH2012,CLLW2015,NHMW2015-1,Z2015,W2015}.
A unitary $t$-design is called \emph{exact} when it simulates all the first $t$ moments of Haar random unitaries and \emph{approximate} when the simulations are with errors.

Traditionally, unitary $t$-designs have been studied for small $t$.
In particular, unitary $2$-designs were intensely studied~\cite{DLT2002,BWV2008a,WBV2008,GAE2007,TGJ2007,DCEL2009,HL2009,DJ2011,BWV2008a,WBV2008,CLLW2015,NHMW2015-1} due to the facts that they are useful in important tasks, such as decoupling~\cite{HaydenTutorial,DBWR2010,SDTR2013,HM14} and randomised benchmarking~\cite{EAZ2005,KLRetc2008,MGE2011,MGE2012}, 
and that the Clifford group is an exact unitary $2$-design~\cite{DLT2002}.
Later, the Clifford group on qubits was also shown to be a unitary $3$-design but not to be a $4$-design~\cite{Z2015,W2015,ZKGD2016}.
For $t \geq 4$, a few applications are known 
(e.g. state discrimination~\cite{S2005}, quantum speed-ups in query complexity~\cite{BH2013} and compressed sensing~\cite{KRT2014,KZD2016}),
but they are of potential importance when strong large deviation bounds are needed.
So far, only a couple of efficient implementations for $t \geq 4$ are known to the best of our knowledge. One is to use a classical tensor product expander and the Fourier transformation, forming approximate unitary $t$-designs for $t=O(N/\log N)$ by using ${\rm poly}(N)$ quantum gates~\cite{HL2009TPE}.  
The other is to use \emph{local random quantum circuits} composed of random two-qubit gates applied onto neighbouring qubits, which achieves approximate unitary $t$-designs for $t = {\rm poly}(N)$ using $O(t^{10} N^2)$ gates~\cite{BHH2012}.

Despite these implementations of unitary designs by quantum circuits, there exists a certain gap between the constructions and physically feasible dynamics in quantum many-body systems. The constructions require a finely structured circuit~\cite{HL2009TPE} or the use of randomly varying interactions~\cite{BHH2012}, while dynamics in physically feasible many-body systems is typically not structured and is generated by a Hamiltonian, which may slightly fluctuate over time but should be based on time-independent one.
Indeed, if we interpret local random quantum circuits on $N$ qubits in terms of Hamiltonian dynamics, the interactions should be changed uniformly at random $O(t^{10} N)$ times before the dynamics achieves unitary $t$-designs. Due to its dependence on the number of particles, the total Hamiltonian should be highly time-dependent and may not be so physically feasible in large systems, resulting in a lack of solid basis of a number of studies of fundamental phenomena in many-body systems based on random dynamics~\cite{PSW2006,GLTZ2006,R2008,GE2016,HP2007,SS2008,S2011,LSHOH2013,S2014,HQRY2016}.
There is also an increasing demand from black hole information science and quantum chaos to fully understand microscopic dynamics of randomisation, where so-called \emph{scrambling} has been intensely studied~\cite{HP2007,SS2008,S2011,LSHOH2013,S2014,HQRY2016,SS2014,RD2015,SS2015}. As scrambling is a weak variant of unitary designs, studying natural Hamiltonians generating unitary designs will bring better understandings in the context.
Further, implementations of unitary designs by Hamiltonian dynamics are of practical importance, helping experimental realisations of designs, as any quantum circuit is fundamentally implemented by engineering Hamiltonians.

In this paper, we provide new constructions of unitary $t$-designs and propose a \emph{design Hamiltonian}, a random Hamiltonian of which dynamics forms a unitary design at \emph{any time} after a threshold time. 
The constructions are based on the scheme of repeating random unitaries diagonal in mutually unbiased bases~\cite{NM2013,NM2014,NHMW2015-1,NHMW2015-2}.
We first show that the process on one qudit achieves unitary $t$-designs after $O(t)$ repetitions if a pair of the two bases satisfies a certain condition, which is met by a pair of any basis and its Fourier basis and that of the Pauli-$X$ and -$Z$ bases.
As the construction works for any space, it will be useful to implement unitary designs in a subspace, such as a symmetric subspace, which is known to demonstrate a quantum advantage in metrology~\cite{OAGKAL2016}.
We then focus on $N$-qubit systems and investigate efficient implementations of random unitaries diagonal in the Pauli-$Z$ basis by quantum circuits. By mapping this problem to a combinatorial problem called a \emph{local permutation check problem}, which can be further reduced to a special type of constrained problems in extremal algebraic theory~\cite{AAK2003,AAK2003-2},
we prove that an approximate unitary $t$-design for $t=o(N^{1/2})$ can be achieved using $O(t N^2)$ gates. In terms of $t$, this drastically improves the previous result~\cite{BHH2012} using $O(t^{10} N^2)$ gates and is essentially optimal.
As higher-designs are useful to improve the performance of applications of lower-designs due to their large deviation bounds~\cite{L2009LDB}, this construction will contribute to improve applications of designs~\cite{D2005,DW2004,GPW2005,ADHW2009,HaydenTutorial,DBWR2010,SDTR2013,HM14,EAZ2005,KLRetc2008,MGE2011,MGE2012,S2005,BH2013,KRT2014,KL15,KZD2016,OAGKAL2016}.
Finally, we introduce design Hamiltonians and present a nearly time-independent one with spin-glass-type interactions, where the interactions need to be varied only $O(t)$ times before the corresponding time-evolution operators form unitary $t$-designs. As a simple consequence, the design Hamiltonians quickly saturate the so-called out-of-time-ordered correlators~\cite{SS2014,RD2015,SS2015} to the Haar averaged values, suggesting a close relation between the design Hamiltonians and quantum chaos.
We also propose a conjecture about the timescale for a natural design Hamiltonian to generate unitary designs, which can be seen as a generalisation of the \emph{fast scrambling conjecture}~\cite{SS2008}.

The paper is organised as follows. In Section~\ref{Sec:pre}, we introduce necessary notation and explain definitions and properties of random unitaries.
All the main results are summarised in Section~\ref{Sec:Main}, of which proofs are provided in Section~\ref{Sec:Proofs}. We conclude and discuss possible future directions in Section~\ref{Sec:Conclusion}. Small lemmas and propositions presented in the paper are proven in Appendices.

\section{Preliminaries} \label{Sec:pre}

We use the following standard asymptotic notation. Let $f(n)$ and $g(n)$ be functions on $\mbb{R}^+$.
We say $f(n) = O(g(n))$ if there exist $c, n_0 >0$ such that $f(n) \leq c g(n)$ for all $n \geq n_0$. 
When there exist $c, n_0 >0$ such that $f(n) \geq c g(n)$ for all $n \geq n_0$, we say $f(n) = \Omega(g(n))$.
If $f(n) = O(g(n))$ and $f(n) = \Omega(g(n))$, we denote it by $f(n) = \Theta(g(n))$.
If $\lim_{n \rightarrow \infty} f(n)/g(n) =0$, we write it by $f(n) = o(g(n))$.
For given $i, j$ ($i<j$), we denote by $[i, j]$ a sequence of numbers from $i$ to $j$, $[i,j]:=\{i, i+1, \cdots, j-1, j \}$.
We also use a floor function $\lfloor x \rfloor$ for $x \in \mathbb{R}$, which is the largest integer less than or equal to $x$.

Let $\mc{H}$ be a Hilbert space and $\mc{B}(\mc{H})$ be a set of bounded operators on $\mc{H}$.
We use several norms of operators and superoperators.
For operators, we use the operator norm $|\!| \cdot |\!|_{\infty}$ and 
the $p$-norm ($p \geq 1$) defined by $|\!| X |\!|_{\infty} := \max_i x_i$, where $\{x_i\}$ are the singular values of $X$, and $|\!| X |\!|_p := (\tr |X|^p)^{1/p}$, respectively.
For a superoperator $\mc{C} : \mc{B}(\mc{H}) \rightarrow \mc{B}(\mc{H})$, we use a family of superoperator norms $|\!| \mc{C} |\!|_{q \rightarrow p}$ ($q,p\geq 1$) and the diamond norm~\cite{KSV2002} defined by
\begin{equation}
|\!| \mc{C} |\!|_{q \rightarrow p} = \sup_{X \neq 0 }\frac{|\!|\mc{C}(X)|\!|_p}{|\!|X|\!|_q}, \hspace{5mm} |\!| \mc{C} |\!|_{\diamond} := \sup_k  |\!|  \mc{C} \otimes {\rm id}_k |\!|_{1 \rightarrow 1},
\end{equation}
respectively, where ${\rm id_k}$ is the identity map acting on a Hilbert space of dimension $k$.

The following are the definitions of Haar random unitaries, random diagonal-unitaries, and unitary $t$-designs.

\begin{Definition}[{\bf Haar random unitaries}]
{\it
Let $\mc{U}(d)$ be a unitary group of degree $d$, and ${\sf H}$ be the Haar measure (i.e. the unique unitarily invariant probability measure) on $\mc{U}(d)$. A \emph{Haar random unitary}
 $U^H$ is a $\mc{U}(d)$-valued random variable distributed according to the Haar measure, $U^H \sim {\sf H}$.}
\end{Definition}

\begin{Definition}[{\bf Random diagonal-unitaries~\cite{NM2013}}]
{\it Let $E=\{ \ket{ k} \}_{k \in [0,d-1]}$ be an orthogonal basis in a Hilbert space $\mc{H}$ with dimension $d$.
Let $\mathcal{D}_E(d)$ be the set of $d \times d$ unitaries diagonal in the basis $E$. Let ${\sf D}_E$ denote a probability measure on $\mathcal{D}_E(d)$ induced by a uniform probability measure on the parameter space $[0,2 \pi)^d$. A \emph{random diagonal-unitary in the basis of $E$}, $D^E$, is a $\mathcal{D}_E(d)$-valued random variable distributed according to ${\sf D}_E$, $D^E \sim {\sf D}_E$.}
\end{Definition}

To define a unitary $t$-design ($t \in \mbb{Z}^+$), let $\mathcal{G}_{U\sim \nu}^{(t)}(X)$ be a superoperator given by 
$\mathcal{G}_{U \sim \nu}^{(t)}(X) := \mathbb{E}_{U\sim \nu} [ U^{\otimes t} X U^{\dagger \otimes t}]$ for any $X \in \mc{B}(\mc{H}^{\otimes t})$, where $\mathbb{E}_{U\sim \nu}$ represents an average over a random unitary $U$ according to a probability measure $\nu$.
An $\epsilon$-approximate unitary $t$-design is then defined as follows.

\begin{Definition}[{\bf An $\epsilon$-approximate unitary $\boldsymbol{t}$-design~\cite{DCEL2009,HL2009}}] \label{Def:Ut}
{\it Let $\nu$ be a probability measure on $\mc{U}(d)$. A random unitary $U \sim \nu$ is an \emph{$\epsilon$-approximate unitary $t$-design} 
 if $|\!|  \mathcal{G}^{(t)}_{U\sim \nu} - \mathcal{G}^{(t)}_{U \sim {\sf H}}  |\! |_{\diamond} \leq \epsilon$.}
\end{Definition}
Due to the property of the diamond norm, unitary $t$-designs are indistinguishable from Haar random unitaries even if we have $t$-copies of the system and are allowed to collectively act on the whole of them.
The following is a trivial but useful lemma about unitary designs.
\begin{Lemma} \label{Lemma:additionalU}
{\it If $U$ is an $\epsilon$-approximate unitary $t$-design, then for any random unitary $V$ independent of $U$, $UV$ and $VU$ are also $\epsilon$-approximate unitary $t$-designs.}
\end{Lemma}
The proof is straightforward and is given in Appendix~\ref{Ap:additionalU}.

We also use the quantum tensor product expander introduced in Ref.~\cite{HH2009}.

\begin{Definition}[Quantum tensor product expander (quantum TPE)~\cite{HH2009}]
{\it Let $\nu$ be a probability measure on $\mc{U}(d)$. Then $\nu$ is a quantum ($\eta$,$t$)-TPE if 
\begin{equation}
|\! | \mathbb{E}_{U \sim \nu}[U^{\otimes t,t}] - \mathbb{E}_{U \sim {\sf H}}[U^{\otimes t,t}] |\! |_{\infty} \leq \eta,
\end{equation}
where $\eta<1$, $U^{\otimes t,t}:=U^{\otimes t} \otimes U^{*\otimes t}$, and $U^*$ is a complex conjugation of $U$.}
\end{Definition}
Note that this definition is equivalent to
\begin{equation}
|\! | \mc{G}_{U \sim \nu}^{(t)} - \mc{G}_{U \sim {\sf H}}^{(t)} |\! |_{2 \rightarrow 2} \leq \eta,
\end{equation}
and hence the difference between a quantum TPE and a unitary $t$-design is just the norm used in their definitions.
The fact that iterating quantum $(\eta,t)$-TPE yields an approximate unitary $t$-design is often used in the literature~\cite{HL2009TPE,BHH2012}, which is formally stated in the following theorem (a proof is given in Appendix~\ref{Sec:ProofTPEtoD} for completeness).

\begin{Theorem} \label{Thm:TPEtoDESIGN}
{\it Let $\nu$ be a quantum ($\eta$,$t$)-TPE. Then iterating the TPE $\ell \geq \frac{1}{\log (1/\eta)} \log \frac{d^{t}}{\epsilon}$ times results in an $\epsilon$-approximate unitary $t$-design.}
\end{Theorem}

\section{Main results} \label{Sec:Main}

Here, we present a summary of our three main results.
We first provide implementations of approximate unitary designs on one qudit in Subsection~\ref{Sec:onequdit}. In Subsection~\ref{Sec:Nqubit}, we consider $N$-qubit systems and show that $\epsilon$-approximate unitary $t$-designs can be implemented by quantum circuits with length $O\bigl(N(t N + \log (1/\epsilon)) \bigr)$. Finally, in Subsection~\ref{Sec:DH}, we propose \emph{design Hamiltonians} and provide a design Hamiltonian with two-body interactions that achieves unitary designs in a short time.

\subsection{One qudit case} \label{Sec:onequdit}

Let us introduce a \emph{Fourier-type pair} of bases, which is important in our result.

\begin{Definition}
{\it A pair of orthogonal bases $(E, F)$ in a $d$-dimensional Hilbert space is called a \emph{Fourier-type pair of bases} if 
each element in $F = \{\ket{\alpha}_F\}_{\alpha \in [0,d-1]}$ is expanded in the basis of $E=\{ \ket{k}_E \}_{k \in [0,d-1]}$ as follows:
\begin{equation}
\ket{\alpha}_F = \frac{1}{\sqrt{d}} \sum_{k \in[0,d-1]} e^{i \theta_{k \alpha}} \ket{k}_E,
\end{equation}
where the phases $\theta_{k \alpha} \in [0, 2 \pi)$ satisfy the condition that
$\forall k,l,\alpha \in [0, d-1]$, $\theta_{k + l,\alpha} = \theta_{k \alpha} + \theta_{l \alpha}$. In the index of $\theta$, $+$ should be an additive operation with respect to which $[0, d-1]$ is an additive group.}
\end{Definition}

The following are two important examples of Fourier-type pairs of bases (see Appendix~\ref{Sec:FourierType} for the proof).

\begin{Lemma} \label{Lemma:FT}
{\it The following pairs of bases are Fourier-type;
\begin{enumerate}
\item any orthogonal basis $\{ \ket{k}\}_{k \in [0,d-1]}$ and its Fourier basis $\{ d^{-1/2} \sum_{k} \omega^{\alpha k} \ket{k}\}_{\alpha \in [0, d-1]}$, where $\omega$ is a $d$th root of unity.
\item the Pauli-$X$ and Pauli-$Z$ bases on $N$ qubits.
\end{enumerate}
}
\end{Lemma}

The former and the later pairs of bases in Lemma~\ref{Lemma:FT} are versions of the position and momentum bases in continuous and discrete spaces, respectively. 
It is known that, if $D^W$ ($W=E,F$) is applied to the state with a large support in the basis of $W$, the resulting state is strongly entangled~\cite{NTM2012, NakataThesis}, implying that $D^W$ has a strong randomisation ability when the initial state is appropriate.
Since each state in one of the mutually unbiased bases has a full support in the other basis, it is natural to expect that alternate applications of $D^E$ and $D^F$ randomise any states and eventually achieve unitary designs. 
Our first main result makes this intuition rigorous. 

\begin{Theorem}{{\bf (Main Result 1)}}~\label{Thm:main}
{\it Let $d = \Omega(t^2 t!^2)$ and $(E, F)$ be a Fourier-type pair of bases.
For independent random diagonal-unitaries $D^E$ and $D'^E$ in the basis of $E$ and $D^F$ in the basis of $F$, $D^E D^F D'^E$ is a quantum $(\eta, t)$-TPE with $\eta$ given by
\begin{equation}
\eta = \frac{(1+t^2)t!^2+t^2}{d} + O\biggl(\frac{t^4  t!^2}{d^{2}} \biggr).
\end{equation}
}
\end{Theorem}

A proof is given in Sec.~\ref{Sec:Pmain}. From Theorems~\ref{Thm:TPEtoDESIGN} and~\ref{Thm:main} and noticing that applying two random diagonal-unitaries in the same basis is equivalent to applying one random diagonal-unitary in that basis, we immediately obtain the following corollary.

\begin{Corollary} \label{Cor:maintDesign}
{\it Let $(E, F)$ be a Fourier-type pair of bases and assume that $d = \Omega(t^2 t!^2)$.
A random unitary $D[\ell]:= D^E_{\ell} D^F_{\ell-1}D^E_{\ell-1} D^F_{\ell}\dots D^E_1 D^F_1 D^E_0$, where $D^E_i$ and $D^F_i$ are independent random diagonal unitaries in the basis of $E$ and $F$, respectively, is an $\epsilon$-approximate unitary $t$-design if 
\begin{equation}
\ell \geq \frac{1}{\log d - 2 \log (t!)} \biggl( t \log d + \log (1/\epsilon) \biggr),
\end{equation}
up to the leading order of $d$ and $t$.}
\end{Corollary}

This construction of designs works for any space, which is not necessarily a whole tensor-product space, and will be useful when we need designs in certain subspaces.
This is the case for instance in quantum metrology, where it was recently shown that almost any random symmetric states are useful to demonstrate a quantum advantage~\cite{OAGKAL2016}. As unitary designs in the symmetric subspace are needed for generating such random states, our construction will help the demonstration of a quantum advantage in metrology.
Another interesting instance is an experimental demonstration of self-thermalisation in isolated quantum many-body systems, which can be done by applying Haar random unitaries or unitary designs onto the system and the environmental system~\cite{PSW2006,GLTZ2006,R2008}. Since the temperature of the system is determined by the total energy in the system and the environment, unitary designs should act on the subspace with restricted energy.
Our construction is suited in this situation because a pair of position and momentum bases of pseudo-particles with fixed energies forms a Fourier-type pair of bases and may be physically feasible to deal with.
Random diagonal-unitaries also have a clear physical interpretation as they are considered to be idealised dynamics by random time-independent Hamiltonians.

Our result should be also compared with the result in Ref.~\cite{HL2009}, where an implementation of approximate unitary $t$-designs was given based on the iterations of classical tensor product expanders and the Fourier transformation. 
The number of iterations in the implementation is approximately $t \log d$. As our result requires approximately only $t$ iterations when $t \ll d$, our construction may seem more efficient. This is however simply a consequence of the fact that random diagonal-unitaries use more randomness than the classical tensor product expander.
We also note that the assumption $d=\Omega(t^2t!^2)$ in Theorem~\ref{Thm:main} and Corollary~\ref{Cor:maintDesign} is for a technical reason. It remains open if $D[\ell]$ constitutes unitary $t$-designs for larger $t$.

\subsection{$N$ qubits case} \label{Sec:Nqubit}

We now focus on $N$-qubit systems.
In particular, we consider applying random diagonal-unitaries in the Pauli-$X$ and -$Z$ bases.
From Corollary~\ref{Cor:maintDesign}, repeating these random diagonal-unitaries yields an $\epsilon$-approximate unitary $t$-design if the number $\ell$ of repetitions satisfies
\begin{equation}
\ell \geq \frac{1}{N - 2 \log_2 ( t!)}\bigl( t N + \log_2 (1/\epsilon) \bigr),
\end{equation}
as long as $2^N = \Omega(t^2t!^2)$.
However, this construction is inefficient because an exact implementation of random diagonal-unitaires by quantum circuits requires an exponential number of local gates.
Thus, we need to find efficient implementations of {\it approximate} random diagonal-unitaries by quantum circuits. 
As the Pauli-$X$ and -$Z$ basese are related by the Hadamard transformation, it suffices to consider those only in the Pauli-$Z$ basis.

We especially study the following family of random diagonal circuits.
Let $\mc{I} = \{ I_i \}$ be a set of $I_i \subset [1,N]$ and denote $M_i:=2^{|I_i|}-1$. At the $i$th step of the circuit, we apply a random diagonal gate ${\rm diag}_Z \{e^{i \varphi_0}, \cdots, e^{i\varphi_{M_i}} \}$ onto the qubits located in $I_i$, where the gate is diagonal in the Pauli-$Z$ basis and the phases $\varphi_k$ ($k \in [0, M_i]$) are chosen independently and uniformly at random from $[0,2\pi)$ every step.
We refer to $|\mc{I}|$ as the length of the circuit. As the circuit is fully specified by $\mc{I}$, we denote it by RDC$(\mc{I})$.

The problem of approximating random diagonal-unitaries in the Pauli-$Z$ basis by RDC$(\mc{I})$ is related to an elementary combinatorial problem, which may be of interest in its own right. We first introduce the combinatorial problem here, and then show the connection to the original problem.

\begin{figure}[tb!]
\centering
\includegraphics[width=140mm, clip]{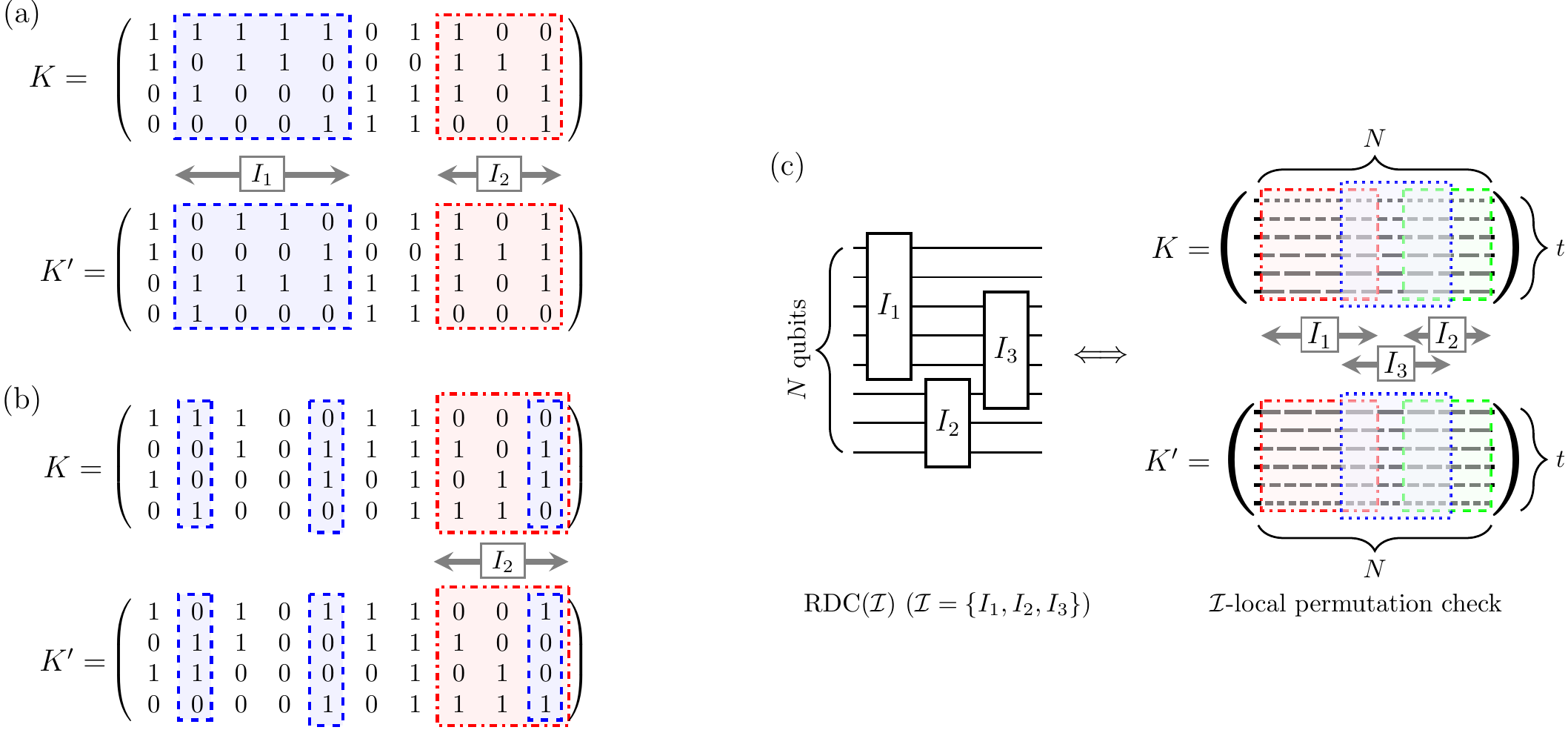}
\caption{
Panel (a) and (b) are examples of local permutation check problems for $t=4$ and $N=10$. 
In Panel (a), $K_{I_1}=\{1111,0110,1000,0001\}$ is a permutation of $K'_{I_1}=\{0110,0001,1111,1000\}$ (blue dashed boxes). However, $K_{I_2}$ is not a permutation of $K'_{I_2}$ (red dash-dotted boxes). Hence, $K$ is an $\{I_1\}$-local but not an $\{I_2\}$-local permutation of $K'$, also implying that $K$ is not a row permutation of $K'$.
In Panel (b), $K$ is identical with $K'$ except blue dashed boxes and is a $2$-local permutation of $K'$.
However, due to the columns in the blue dashed boxes, $K$ fails to be a $3$-local permutation of $K'$. To observe this, compare e.g. $K_{I_2}$ and $K'_{I_2}$ in red dash-dotted boxes. 
Panel (c) illustrates a relation between RDC$(\mc{I})$ and an $\mc{I}$-local permutation check problem.
As diagonal gates acts on $I_1$, $I_2$ and $I_3$, we first check if $K$ is a $\{I_1, I_2, I_3\}$-local permutation of $K'$. That is, we check the permutation relations between sets of rows in the red dash-dotted, green dotted, and blue dashed boxes. If $K$ is $\{I_1, I_2, I_3 \}$-local but not a row permutation of $K'$, then $\bra{K, K'} \mbb{E}_{D ^Z\sim {\sf RDC}(\mc{I})}[(D^Z)^{\otimes t,t}] \ket{K, K'}=1$ and otherwise $0$.
}
\label{Fig:LPC}
\end{figure}

Let $K$ and $K'$ be $t \times N$ matrices with elements in $\{0,1 \}$.
For given $s \in [1,t]$ and $I \subset [1,N]$, we denote a subsequence $( K_{s, m} )_{m \in I}$ of the $s$th row of $K$ by $K_{s, I}$, and a set $\{K_{s, I}\}_{s \in [1,t]}$ of such subsequences over all $s$ by $K_I$.
We use the same notation also for $K'$.
Let $\Omega$ be a canonical map that rearranges the subsequences $K_I$ in ascending order, where the subsequences are regarded as binary numbers.
For $\mc{I} = \{ I \}$, we say that $K$ is an \emph{$\mc{I}$-local permutation} of $K'$ if $\forall I \in \mc{I}$, $\Omega(K_I) =  \Omega(K'_I)$. 
In particular, we say $K$ is a \emph{row permutation} of $K'$ if $\Omega(K_I) =  \Omega(K'_I)$ for $I=[1,N]$,
which simply implies that a set of rows of $K$ is a permutation of that of $K'$.
In the following, we denote by $\mc{I}_r$ a set of all subsets in $[1,N]$ with $r$ elements.
Using this notation, we define \emph{local permutation check problems} as follows.

\begin{Definition}[{\bf Local permutation check problems}]
{\it 
Let $K$ and $K'$ be $t \times N$ matrices with elements in $\{0,1 \}$.
For a given $\mc{I} = \{ I_i \}$ ($I_i \subset [1,N]$), the task of the \emph{$\mc{I}$-local permutation check problem} is to count the number of pairs $(K,K')$ such that $K$ is not a row permutation but an $\mc{I}$-local permutation of $K'$. We denote the number of such pairs by $\Lambda(\mc{I})$.
In particular, for $\mc{I}_r$,
we call the problem an \emph{$r$-local permutation check problem} and denote the number of pairs by $\Lambda_r$.
}
\end{Definition}

\begin{figure}[tb!]
\centering
 \includegraphics[width=120mm, clip]{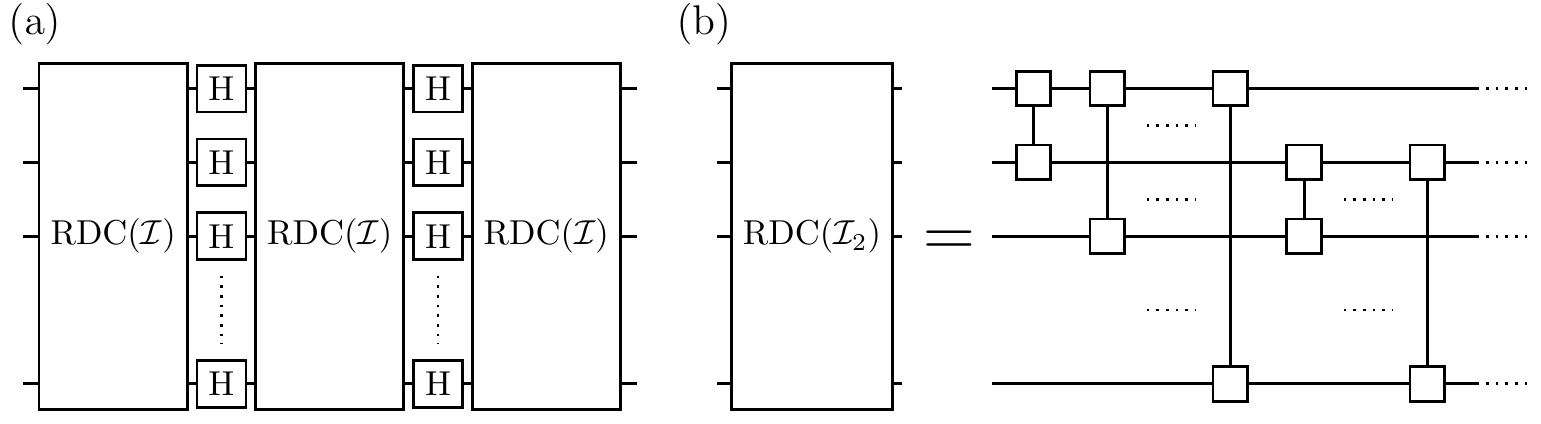}
\caption{Panel (a) depicts iterations of RDC($\mc{I}$) and the Hadamard transformation. Panel (b) shows RDC($\mc{I}_2)$, where random diagonal two-qubit gates are applied onto all pairs.
The circuit is called RDC$_{\rm disc}^{(t)}(\mc{I}_2)$, when each two-qubit gate is replaced with $({\rm diag}_Z\{1, e^{i \varphi_1} \} \otimes {\rm diag}_Z\{1, e^{i \varphi_2} \} )
{\rm diag}_Z\{1, 1,1, e^{i \vartheta} \}$, where the phases $\phi_1, \phi_2$ and $\vartheta$ are chosen from discrete sets given in the main text.}

\label{Fig:DUtQC}
\end{figure}

For a couple of examples of local permutation checks, see Fig.~\ref{Fig:LPC}.
The following lemma provides the connection between
the $\mc{I}$-local permutation check problem and implementations of quantum TPEs by RDC$(\mc{I})$ (see Appendix~\ref{Sec:PLP} for the proof).
\begin{Lemma} \label{Lemma:LP}
{\it
Let $2^N=\Omega(t^2 t!^2)$.
For a given $\mc{I}=\{ I_i \}$ where $I_i \subset [1,N]$,
iterating RDC$(\mc{I})$ and the Hadamard transformation $H_N$ on $N$ qubits, 
such as RDC$(\mc{I}) H_N RDC(\mc{I}) H_N RDC(\mc{I})$ (see Fig.~\ref{Fig:DUtQC} (a)), yields a quantum $(\tilde{\eta}, t)$-TPE where
\begin{equation}
\tilde{\eta} \leq \eta + 3 t! \frac{\Lambda (\mc{I})}{2^{tN}} + \biggl(\frac{\Lambda (\mc{I})}{2^{tN}} \biggr)^2
\end{equation}
with $\eta = \frac{(1+t^2) t!^2+t^2 }{2^N} + O(\frac{t^4 t!^2 }{2^{2N}})$.
}
\end{Lemma}

To obtain our second main result, RDC$(\mc{I}_2)$ (see Fig.~\ref{Fig:DUtQC} (b)) and the $2$-local permutation check problem are of particular importance.
Due to the result in Ref.~\cite{NKM2014}, we know that $\Lambda_2 = 0$ for $t \leq 3$. 
When $t \geq 4$, the problem can be rephrased as an extremal problem under dimension constraints, which is a constrained problem in extremal algebraic theory~\cite{AAK2003,AAK2003-2}.
By solving the problem, we obtain the following key lemma (see Sec.~\ref{Sec.PPairs} for the proof).

\begin{Lemma} \label{Lemma:PairsLP}
{\it 
For the $2$-local permutation check problem, it holds that $\Lambda_2 \leq 2^{2t^2 + (t-1) N}$.
}
\end{Lemma}

It immediately follows from Lemmas~\ref{Lemma:LP} and~\ref{Lemma:PairsLP} that, when $t=o(N^{1/2})$, iterating RDC$(\mc{I}_2)$ and the Hadamard transformation is a quantum $(\tilde{\eta},t)$-TPE, where 
\begin{equation}
\tilde{\eta} \leq 2^{2t^2+2-N}t! + O\bigl(t^2 t!^2  2^{-N} \bigr), \label{Eq:tildeeta}
\end{equation}
from which we obtain an efficient implementation of a unitary $t$-design due to Theorem~\ref{Thm:TPEtoDESIGN}.
We can further reduce the number of randomness in the implementation by replacing all gates in RDC$(\mc{I}_2)$ with those in the form of 
\begin{equation}
\bigl({\rm diag}\{1, e^{i \varphi_1} \} \otimes {\rm diag}\{1, e^{i \varphi_2} \} \bigr)
{\rm diag}\{1, 1,1, e^{i \vartheta} \}. \label{Eq:RDCno;j4}
\end{equation}
When $\varphi_1$ and $\varphi_2$ are chosen independently from $\{ 2 \pi m/a : m \in [0,a-1] \}$ uniformly at random, and $\vartheta$ is chosen from $\{ 2 \pi m/b : m \in [0,b-1] \}$, we denote the circuit RDC$_{\rm disc}(\mc{I}_2:a,b)$.
Using the same technique as in Ref.~\cite{NKM2014}, we obtain the following lemma.
\begin{Lemma} \label{Lemma:last}
Let $a \geq t+1$ and $b\geq \lfloor t/2 \rfloor+1$. Then, we have
$\mathbb{E}[{\rm RDC}_{\rm disc}(\mc{I}_2:a,b)^{\otimes t,t}] = \mathbb{E}[{\rm RDC}(\mc{I}_2)^{\otimes t,t}]$.
\end{Lemma}
In particular, we denote RDC$_{\rm disc}(\mc{I}_2:t+1, \lfloor t/2 \rfloor+1)$ simply by RDC$_{\rm disc}^{(t)}(\mc{I}_2)$, where one two-qubit gate requires $2 \log_2(t+1) + \log_2 (\lfloor t/2 \rfloor+1) < 3 \log_2(t+1)$ random bits.
Together with all of these, we obtain our second main result.

\begin{Theorem}{{\bf (Main Result 2)}} \label{Thm:2}
{\it 
Let $t = o(N^{1/2})$.
Then, iterating RDC$_{\rm disc}^{\ (t)}(\mc{I}_2)$ and the Hadamard transformation on $N$ qubits such as 
$\bigl(RDC_{\rm disc}^{(t)}(\mc{I}_2) H_N \bigr)^{2\ell} RDC_{\rm disc}^{(t)}(\mc{I}_2)$ yields an $\epsilon$-approximate unitary $t$-design if
\begin{equation}
\ell \geq t + \frac{ \log_2 (1/\epsilon)}{N},
\end{equation}
up to the leading order of $N$ and $t$.
The total number of two-qubit gates and random bits are given by
\begin{align}
&\text{ \# of two-qubit gates } = \Theta\bigl( N \bigl( t N +\log_2 (1/\epsilon) \bigr)\bigr),\\
&\text{ \# of random bits } = \Theta\bigl((\log_2 t)N \bigl( t N +\log_2 (1/\epsilon) \bigr) \bigr),
\end{align}
respectively.
}
\end{Theorem}

\begin{table}[t!]
\centering
\begin{tabular}{l| |c|c|c} 
 & Total number of gates & $t$  & Non-commuting depth~\cite{NHMW2015-2} \\ \hline \hline
Classical tensor expanders~\cite{HL2009TPE} & poly$(N)$ & $O(N/\log N))$  & poly$(N)$    \\ \hline
Local random circuits~\cite{BHH2012} & $O\bigl(t^9N(tN + \log (1/\epsilon))\bigr)$  & poly$(N)$ & $O\bigl(t^9(tN + \log (1/\epsilon))\bigr)$  \\ \hline
Random diagonal circuits & $O\bigl(N(tN + \log_2 (1/\epsilon))\bigr)$  & $o(N^{1/2})$ & $O\bigl(t + \frac{1}{N} \log_2 (1/\epsilon) \bigr)$  
\end{tabular}
\caption{A comparison between quantum circuit constructions of unitary $t$-designs on $N$ qubits, which works for $t \geq 3$.
The total number of quantum gates to achieve classical tensor expanders is known to be poly$(N)$, but is not explicitly presented in Ref.~\cite{HL2009TPE}. The non-commuting depth was introduced in Ref.~\cite{NHMW2015-2} and is defined by the circuit depth when each commuting part of the circuit is counted as one step. The non-commuting depth may be of experimental importance.}\label{Tab:Designcomparison}
\end{table}

Although we assume in Theorem~\ref{Thm:2} that $t = o(N^{1/2})$. However, we believe that Theorem~\ref{Thm:2} holds even for $t=o(N/\log N)$, which comes from the conjecture explained in more detail in Sec.~\ref{Sec.PPairs}.

In terms of $t$, Theorem~\ref{Thm:2} drastically improves the previous result using $O\bigl(t^9 N(t N + \log (1/\epsilon))\bigr)$ two-qubit gates~\cite{BHH2012} (see also Table~\ref{Tab:Designcomparison} for the comparison) and is essentially optimal when the design is defined on a finite set of unitaries.
This is because the support of a unitary $t$-design should contain at least $O(2^{2tN})$ unitaries~\cite{RS2009}. 
Thus, when each gate in a random quantum circuit is chosen from a finite set, the scaling of the length necessary for the circuit achieving a $t$-design cannot be substantially better than linear in $t$.
 
In practical uses of unitary designs such as decoupling~\cite{HaydenTutorial,DBWR2010,SDTR2013,HM14} and randomised benchmarking~\cite{EAZ2005,KLRetc2008,MGE2011,MGE2012}, unitary $2$-designs are known to be sufficient, which can be achieved more efficiently than our construction if one uses a Clifford circuit~\cite{CLLW2015}. However, unitary $4$-designs are needed in a few applications~\cite{S2005,BH2013,KRT2014}, which cannot be achieved by any Clifford circuit~\cite{Z2015}.
Moreover, higher-designs are generally more useful than lower-designs because they have stronger large deviation bounds~\cite{L2009LDB}, which are finite approximations of the concentration of measure for Haar random unitaries stating that values of any slowly varying function on a unitary group are likely to be almost constant if the dimension is large~\cite{L2001}.
This implies that using higher-designs in any applications of unitary designs results in better performance.
As our implementation provides a shorter quantum circuit for $t$-designs than the existing ones~\cite{HL2009,BHH2012},
it contributes to improve the performance of quantum protocols using unitary designs~\cite{D2005,DW2004,GPW2005,ADHW2009,HaydenTutorial,DBWR2010,SDTR2013,HM14,EAZ2005,KLRetc2008,MGE2011,MGE2012,S2005,BH2013,KRT2014,KL15,OAGKAL2016}.

This construction of approximate designs also has advantages from an experimental point of view.
As highlighted in Ref.~\cite{NHMW2015-1,NHMW2015-2}, the quantum circuits repeating RDC$(\mc{I}_2)$ or RDC$_{\rm disc}^{(t)}(\mc{I}_2)$ and the Hadamard transformation are divided into a constant number of commuting parts. Indeed, only non-commuting parts are the Hadamard parts. 
Because the gates in each commuting part do not have any temporal order, they can be applied simultaneously in experimental realisations, making the implementations more robust.
Hence, the commuting structure of our construction may help reducing the practical time and increasing the robustness of the implementations. This property can be rephrased in terms of the \emph{non-commuting depth} proposed in Ref.~\cite{NHMW2015-2} (see Table~\ref{Tab:Designcomparison}).

\subsection{Hamiltonian dynamics and unitary designs} \label{Sec:DH}

In the last decade, unitary designs were revealed to be the key to understanding fascinating phenomena in complex quantum many-body systems~\cite{PSW2006,GLTZ2006,R2008,GE2016,HP2007,SS2008,S2011,LSHOH2013,S2014,HQRY2016}, in most of which the dynamics is assumed to be so random that it can be described by unitary designs. This assumption may be reasonable as a first approximation.
However, due to the lack of full understanding of natural microscopic dynamics generating unitary designs, it is not clear to what extent the assumption can be justified. Most recently, the idea of \emph{scrambling} was introduced in black hole information science~\cite{HP2007,SS2008}.  The main concern there is the \emph{fast scrambling conjecture}, stating that the shortest time necessary for natural dynamics to scramble many-body systems scales logarithmically with the system size~\cite{SS2008,S2011,LSHOH2013,S2014,HQRY2016}. 
While it is known that $0$-dimensional systems, where all particles interact with each other, can be scrambled in a constant time~\cite{M2016}, the conjecture is strongly believed to hold in higher dimensions.
The fast scrambling conjecture originally arises from a thought experiment concerning the black hole evaporation and the no-cloning theorem~\cite{SS2008}, but has been also studied intensely in connection with quantum chaos~\cite{SS2014,RD2015,SS2015}. 
So far, several inequivalent definitions of scrambling were proposed~\cite{SS2008,LSHOH2013,HQRY2016}. Although they are useful for clarifying the relations between scrambling and other notions of randomisation such as unitary designs and the OTO correlators diagnosing quantum chaos~\cite{SS2014,RD2015,SS2015}, there does not seem to be consensus on a rigorous mathematical definition of scrambling.

Here, we introduce \emph{design Hamiltonians} as a unifying framework for studying randomising operations by physically natural Hamiltonian dynamics. In terms of the design Hamiltonians, we generalise the fast scrambling conjecture and propose a \emph{natural design Hamiltonian conjecture}.
We then construct a design Hamiltonian, where the interactions need to be changed only a few times before the corresponding time-evolution operators form unitary designs. This is in sharp contrast to the Hamiltonian dynamics based on local random quantum circuits~\cite{BHH2012}, which will be elaborated on later.

We start with the definition of \emph{$k$-local Hamiltonians}.

\begin{Definition}[$k$-local Hamiltonians~\cite{KSV2002}]
{\it 
Let $\Lambda_j \subset [1,N]$ such that $|\Lambda_j| \leq k$ and $\Lambda_i \neq \Lambda_j$ if $i\neq j$. A $k$-local Hamiltonian $H$ on $N$ qubits is one in the form of 
$H= \sum_{i} H_i$, where each term $H_i$ may depend on time, acts non-trivially only on the qubits in $\Lambda_i$ and satisfies $|\!| H_i |\!|_{\infty} \leq 1$. We denote a set of all $k$-local Hamiltonians by $\mathfrak{H}_k$.}
\end{Definition}

The interactions in $k$-local Hamiltonians are not necessarily geometrically local on lattice systems. They are rather interpreted as interactions on a given graph, where each vertex represents a particle. To normalise the time scale of the dynamics, we also assumed that the strength of each local interaction is bounded.
In the following, to avoid confusion, we always use small $t$ and capital $T$ for $t$-designs and time, respectively. Denoting by $U_H(T) := \mc{T}\exp[-i \int_0^T ds H(s)]$, where $\mathcal{T}\exp$ is the time-ordered exponential, the time evolution operator at time $T$ generated by a possibly time-dependent Hamiltonian $H$, we now introduce a \emph{$t$-design Hamiltonian with $k$-local interactions} as follows.

\begin{Definition}[An $\epsilon$-approximate $t$-design Hamiltonian with $k$-local interactions] \label{Def:DesHam}
{\it Let $\mathfrak{H}^{(t)}_{\epsilon} \subset \mathfrak{H}_k$ and ${\sf Ham}^{(t)}_{\epsilon}$ be a probability measure on $\mathfrak{H}^{(t)}_{\epsilon}$. If there exists $T_0>0$ such that, $\forall T \geq T_0$, 
a random unitary $U_H(T)$ generated by $H \sim {\sf Ham}^{(t)}_{\epsilon}$
is an $\epsilon$-approximate unitary $t$-design,
the random Hamiltonian $H$ is called an \emph{$\epsilon$-approximate $t$-design Hamiltonian with $k$-local interactions}.
We also call the shortest such time $T_0$ a \emph{design time of $H$} .
}
\end{Definition}

\begin{figure}[tb!]
\centering
\includegraphics[width=160mm, clip]{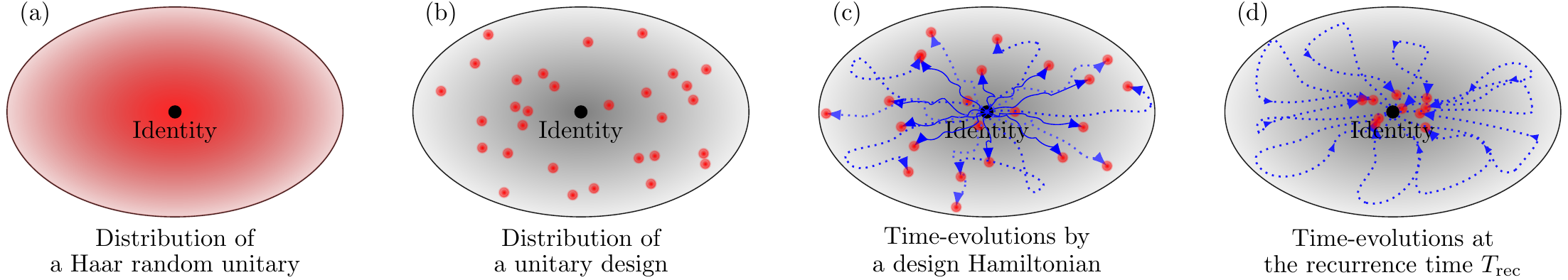}
\caption{
Schematic figures illustrating the distributions of random unitaries in a whole unitary group. For the visualisation, the unitary group is represented by an ellipse and each red dot corresponds to a unitary operator. Panel (a) illustrates a Haar random unitary, which is uniformly and continuously distributed over the whole unitary group. For unitary designs, the distribution is not necessarily continuous and is often defined on a finite support, which is depicted in Panel (b).
Panel (c) provides an intuitive picture of time-evolution operators generated by a design Hamiltonian, starting from the identity. As time passes, a design Hamiltonian generates random unitary distributed over the whole unitary. The time evolution is illustrated by a trajectory in the panel. When the design Hamiltonian is defined on a finite ensemble of Hamiltonians, there exists a time $T_{\rm rec}$, where all time evolution operators are in the neighbourhood of the identity, due to the Poincar\'{e} recurrence theorem as depicted in Panel (d).
}
\label{Fig:DesHamOO}
\end{figure}

Note that, in this sense, there is no design Hamiltonian on a finite ensemble of time-independent Hamiltonians.
Due to the Poincar\'{e} recurrence theorem~\cite{BL1957}, the time-evolution operator generated by a time-independent Hamiltonian is in the neighbourhood of the identity operator at the recurrence time. Although the time-evolution operators generated by other Hamiltonians are possibly not the identity at the recurrence time of one Hamiltonian, we can always find the time $T_{\rm rec}$ where all operators are close to the identity. Hence, at that time, an ensemble of time-evolution operators does not form unitary designs (see also Fig.~\ref{Fig:DesHamOO}).
However, this problem can be avoided if we consider time-dependent Hamiltonians or a continuous ensemble of time-independent Hamiltonians. We can also relax the condition of $\forall T \geq T_0$ to most of the time after $T_0$. For simplicity, in this paper, we define the design Hamiltonian as in Definition~\ref{Def:DesHam}.

As our main purpose is to find physically natural Hamiltonians generating unitary designs, we are most interested in the design Hamiltonians which are not finely structured, are time-independent and are with geometrically local interactions. In addition, we may further require that, due to the fast scrambling conjecture, the design time scales logarithmically with the system size, which may depend on $t$. Thus, we arrive at the following conjecture. 

\begin{Conjecture}[Natural design Hamiltonian conjecture]
{\it
There exist $\epsilon$-approximate $t$-design Hamiltonians on $N$ qubits that satisfy the following three conditions:
\begin{enumerate}\vspace{-2mm}
\item the interactions are geometrically local, \vspace{-2mm}
\item the interactions are all time-independent,\vspace{-2mm}
\item the design time is given by $O(t \log N)$, which may also depend on $\epsilon$.\vspace{-2mm}
\end{enumerate}
}
\end{Conjecture}

In general, the Hamiltonians with random interactions are expected to exhibit many-body localization~\cite{SPA2013,HNO2014, NH2015}, preventing the corresponding dynamics from achieving unitary designs quickly. 
However, this is not always the case. 
For instance, the dynamics of a Majorana fermion model with random four-body interactions, also known as the Sachdev-Ye-Kitaev (YSK) model~\cite{SY1993,K2015} is known to be strongly chaotic~\cite{K2015Feb,MSS2016} and is likely to achieve unitary designs at least on the low energy subspace. Although the SYK model consists of all-to-all interactions and does not meet the first condition of the conjecture, the further investigation of the model may help the search of natural design Hamiltonians satisfying all the three conditions.

The conjecture is based on an established language of unitary designs and so will be helpful to explore randomising operations in physically natural systems in a mathematically rigorous manner.
We note that the conjecture is not only of theoretical interest but also of practical importance because, by applying such a random Hamiltonian onto a system, a unitary design will be spontaneously obtained. Most importantly, there is no need to change the interactions and no fine control of time is required.
This will drastically simplify the implementations of unitary designs in experiments, also resulting in the simplification of many quantum protocols~\cite{D2005,DW2004,GPW2005,ADHW2009,HaydenTutorial,DBWR2010,SDTR2013,HM14,EAZ2005,KLRetc2008,MGE2011,MGE2012,S2005,BH2013,KRT2014,KL15,OAGKAL2016}.

The construction of designs by local random quantum circuits~\cite{BHH2012} can be naturally translated into design Hamiltonians: \emph{a random Hamiltonian with neighbouring two-body interactions is a $t$-design Hamiltonian if the interactions vary randomly and independently at every time step}. 
Such varying interactions can be considered to be fluctuations induced by white noise on two-body interactions~\cite{OBKBWE2016}.
This design Hamiltonian $H_{\rm rand}$ satisfies the first condition of the conjecture, as it uses only neighbouring interactions, but not the second and the third ones. Indeed, 
to achieve a unitary $t$-design by the dynamics of $H_{\rm rand}$, the interactions should be changed $O(t^{10}N)$ times uniformly at random. This is far from time-independent and takes much longer than $O(t \log N)$. 

Here, we are more concerned with the second condition of the conjecture and provide a design Hamiltonian $H_{XZ}$ based on Theorem~\ref{Thm:2}.
We start with introducing a parameter set $\mc{P}_t(c)$ by
\begin{align}
\mc{P}_t(c) &= \biggl\{ \frac{m}{2 (\lfloor t/2 \rfloor + 1)}: m \in \bigl[-c,c \bigr] \biggr\}. \label{Eq:parameter}
\end{align}
We then define finite sets of commuting Hamiltonians:
\begin{align}
\mathfrak{H}_Z^{(t)}
&:=
\biggl\{-\sum_{j<k} J_{ik} Z_j \otimes Z_k - \sum_{j} B_j Z_j \biggr\}_{J_{jk} \in \mc{P}_t(J), B_{j} \in \mc{P}_t(B)}, \label{HamX}\\
\mathfrak{H}_X^{(t)}
&:=
\biggl\{-\sum_{j<k} J_{ik} X_j \otimes X_k - \sum_{j} B_j X_j \biggr\}_{J_{jk} \in \mc{P}_t(J), B_{j} \in \mc{P}_t(B)}, \label{HamZ}
\end{align}
where $J=\frac{\lfloor t/2 \rfloor}{2}$ and $B=\lfloor t/2 \rfloor+\frac{1}{2}$.
These types of disordered Hamiltonians are similar to those in many-body localised systems~\cite{SPA2013,HNO2014, NH2015}, while interactions typically decay with increasing distance in such systems.
Finally, we introduce a notation $\in_{\rm R}$ which implies that the left-hand side is drawn uniformly at random from the set in the right-hand side.

\begin{figure}[tb!]
\centering
\includegraphics[width=160mm,clip]{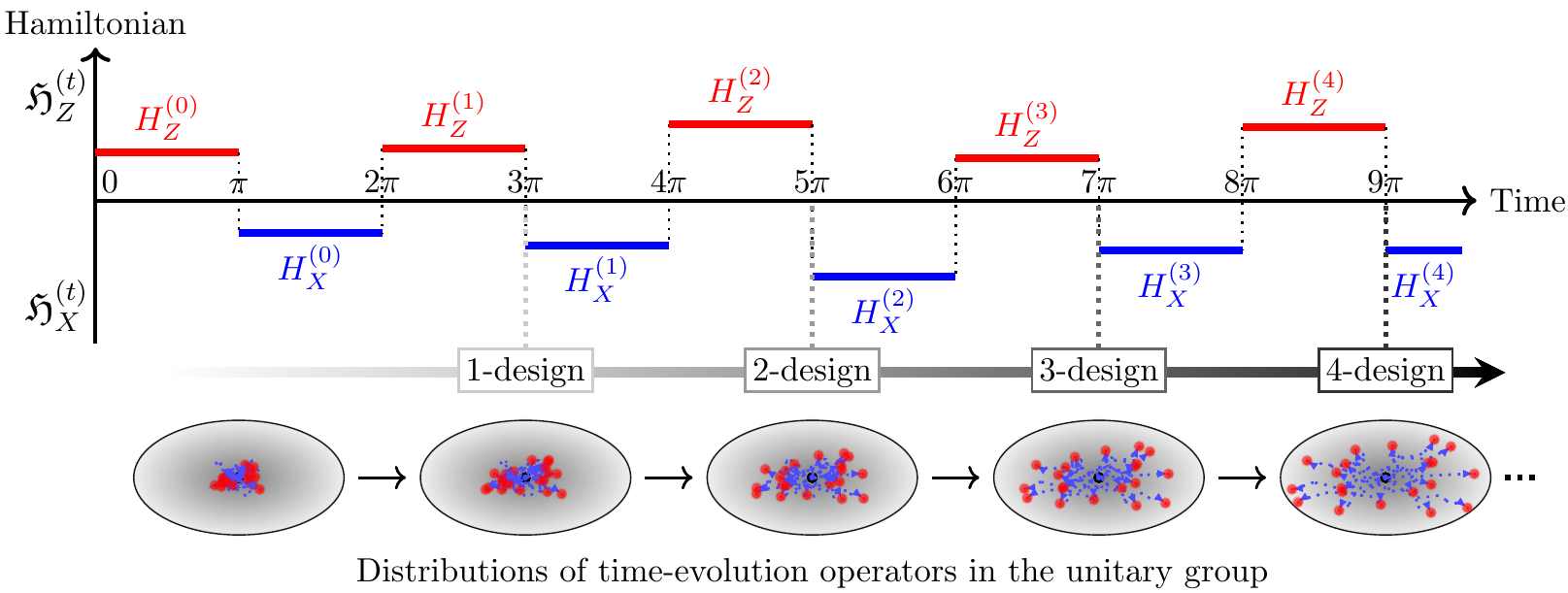}
\caption{A schematic figure about the design Hamiltonian $H_{XZ} \in_{\rm R} \mathfrak{H}_{XZ}^{(t)}$. At each time interval $m$, $H_Z^{(m)}$ or $H_X^{(m)}$ is chosen uniformly at random from $\mathfrak{H}_{Z}^{(t)}$ or $\mathfrak{H}_{X}^{(t)}$, respectively. As depicted at the bottom of the figure, a random unitary generated by $H_{XZ}$ rapidly spreads over the whole unitary group and forms unitary designs in a short time independent of the system size.}
\label{Fig:DesHam}
\end{figure}

Our third result is given as follows (see Sec.~\ref{Sec:HamDes} for the proof).
\begin{Corollary}[Main result 3] \label{Cor:3}
{\it
Let $t = o(N^{1/2})$ and $\mathfrak{H}_{XZ}^{(t)}$ be a set of $2$-local time-dependent Hamiltonians in the form of
\begin{equation}
H_{XZ}(T) = \begin{cases}
H_Z^{(m)} & \text{if\ \  $ 2m \pi \leq T < (2m+1) \pi$},\\
H_X^{(m)} & \text{if\ \  $  (2m+1) \pi \leq  T < 2(m+1) \pi$},
\end{cases} \label{Eq:designHam}
\end{equation}
where $T$ denote time, and $H_W^{(m)} \in  \mathfrak{H}_W^{(t)}$ for any $m =0,1,\cdots$ ($W=X,Z$). Then, the random Hamiltonian $H_{XZ} \in_{\rm R} \mathfrak{H}_{XZ}^{(t)}$ is an $\epsilon$-approximate $t$-design Hamiltonian. The design time of $H_{XZ}$ is at most $ \bigl(2t+1 + \frac{2}{N} \log_2 (1/\epsilon) \bigr)\pi$.
}
\end{Corollary}

Since $H_{XZ}$ is composed of $H_Z$ and $H_X$, both of which exhibit many-body localization, one may think that the time-evolution operators generated by $H_{XZ} \in_{\rm R} \mathfrak{H}_{XZ}^{(t)}$ shall not spread over the whole unitary group. However, due to the periodic change of the interaction basis, the localization indeed helps the time-evolution operators to be uniform.
This can be observed from the fact that a random unitary diagonal in a fixed basis has a strong randomisation power when the initial state has a large support in that basis~\cite{NTM2012, NakataThesis}. Since a localized state in one basis has a large support in the complementary basis, the time-evolution by $H_Z$ ($H_X$) randomises the localized eigenstates of $H_X$ ($H_Z$) strongly. For this reason, it is natural to expect that the time-evolution operators generated by $H_{XZ}$ eventually form a unitary design, as rigorously proven in Corollary~\ref{Cor:3}.

Note that our specific choice of the parameters in the Hamiltonians $H_Z$ and $H_X$, namely $J_{jk} \in_{\rm R} \mc{P}_t(J)$ and $B_{j} \in_{\rm R} \mc{P}_t(B)$, is to minimize the randomness needed to construct a design Hamiltonian. It is possible to choose the parameters from different sets as long as they are sufficiently random, where the design time will be accordingly changed. From a physical point of view, it may be interesting to consider physically feasible noises as parameter sets, which is in the same spirit as Ref.~\cite{OBKBWE2016}.

Corollary~\ref{Cor:3} implies that the random Hamiltonian $H_{XZ}$ quickly generates the time evolution which can be hardly distinguished from completely random one (see also Fig.~\ref{Fig:DesHam}). Most notably, the design time is $O(t)$ and independent of the system size.
As a simple consequence, any correlation functions at time $T$ in the system described by such a Hamiltonian quickly converges to the Haar averaged values. One of the important instances is the $2t$-point OTO correlator, which is expected to diagnose quantum chaos and has been studied in strongly correlated systems~\cite{SS2014,RD2015,SS2015}. As the $2t$-point OTO correlators are polynomials of a unitary with degree $t$, their values in the system of a random Hamiltonian $H_{XZ}$ are $\epsilon$-close to the Haar random averages when $T \gtrsim (2t +1+ \frac{2}{N} \log_2 (1/\epsilon))\pi$. Furthermore, due to the large deviation bounds for unitary designs~\cite{L2009LDB}, this implies that almost any Hamiltonian in $\mathfrak{H}_{XZ}^{(t)}$ saturates the $2t$-point OTO correlators to the Haar random averages in a short time irrespective of the system size.
As the OTO correlators are saturated in quantum chaotic systems~\cite{HQRY2016}, our result indicates a close connection between the Hamiltonians in $\mathfrak{H}_{XZ}^{(t)}$ and quantum chaos, which suggests that the framework of design Hamiltonians may be useful to investigate the dynamics in quantum chaotic systems. This is also supported by a recently clarified relation between unitary designs and quantum chaos~\cite{RB2016}.

\begin{table}[t!]
\centering
\begin{tabular}{c| |c|c|c} 
Design Hamiltonian & Interactions & Time-dependence  & Design time \\ \hline \hline
$H_{\rm rand}$ & Nearest neighbour interactions& Highly dependent  & $O(t^{10} N)$    \\ \hline
$H_{XZ}$ & All-to-all two-body interactions  & Nearly time-independent & $O(t)$  \\
\end{tabular}
\caption{A comparison of design Hamiltonians, $H_{\rm rand}$~\cite{BHH2012} and $H_{XZ}$, in terms of the three conditions of the natural design Hamiltonian conjecture. 
The design time of $H_{XZ}$ is much shorter than that of $H_{\rm rand}$ both in terms of $t$ and $N$. Although the improvement in terms of $t$ is generic to $H_{XZ}$, that in terms of $N$ is possibly due to its all-to-all interactions (see the main text).}\label{Tab:comparison}
\end{table}

In Table~\ref{Tab:comparison}, we compare two design Hamiltonians $H_{\rm rand}$ and $H_{XZ}$.
We emphasise that the design time $O(t)$ of $H_{XZ}$ is significantly faster than the design time $O(t^{10} N)$ of $H_{\rm rand}$ in terms of both $t$ and $N$. 
We should note, however, that although the improvement in terms of $t$ is intrinsic to $H_{XZ}$, the improvement in terms of $N$ may be rather due to the all-to-all interactions of $H_{XZ}$. 
Such interactions may naturally appear in cavity QED~\cite{SM2002,GLG2011,SS2011} due to the cavity modes mediating long-range interactions, and unitary designs may possibly be realised in a short time. 
Nevertheless, for the fair comparison with $H_{\rm rand}$, the realisation of all-to-all interactions by neighbouring ones should be taken into account. This can be achieved if every particle travels all corners of the system and interacts with all the other particles, taking $O(N)$ time. Hence, when the interactions are neighbouring, the actual time for $H_{XZ}$ to achieve unitary designs is considered to be $O(tN)$, also implying that it does not violate the fast scrambling conjecture.

Unfortunately, both design Hamiltonians $H_{\rm rand}$ and $H_{XZ}$ do not satisfy all three conditions of the natural design Hamiltonian conjecture.
However, we believe that existence of two design Hamiltonians $H_{\rm rand}$ and $H_{XZ}$, and previous analyses on the original fast scrambling conjecture~\cite{SS2008,S2011,LSHOH2013,S2014,HQRY2016} provide substantial evidences for the natural design Hamiltonian conjecture.

\section{Proofs} \label{Sec:Proofs}
In this section, we provide proofs of theorems and lemmas given in Section~\ref{Sec:Main}. We first introduce additional notation and useful lemmas in Subsection~\ref{SS:NandL}.
The proof of our first main result, Theorem~\ref{Thm:main}, is given in Subsection~\ref{Sec:Pmain}. We prove Lemma~\ref{Lemma:PairsLP} in Subsection~\ref{Sec.PPairs}, which is the key lemma to obtain our second main result, and conclude this section by showing Corollary~\ref{Cor:3} about design Hamiltonians in Subsection~\ref{Sec:HamDes}.

\subsection{Additional notation and lemmas} \label{SS:NandL}

Let $E=\{ \ket{k}_E \}_{k \in [0,d-1] }$ and $F=\{ \ket{\alpha}_F \}_{\alpha \in [0,d-1] }$ be orthogonal bases in $\mc{H}$. 
As we deal with $t$ copies of the Hilbert space, $\mc{H}^{\otimes t}$, 
we denote $[0,d-1]^t$ by $\mc{N}$ and
introduce bases $\{ \ket{ \mbf{k}}_W \}_{\mbf{k} \in \mc{N}}$ ($W=E,F$) in $\mc{H}^{\otimes t}$, where $\ket{ \mbf{k}}_W = \bigotimes_{s=1}^t \ket{ {k_s}}_W$, $\mbf{k}=(k_1, \cdots, k_t)^T \in \mc{N}$, and $T$ represents the transpose. 
In the following, we always label the basis $E$ and $F$ by Latin and Greek alphabets, respectively, and do not write the subscript $E$ and $F$ explicitly.

Let $S_t$ be a permutation group of degree $t$.
For $\pi \in S_t$, we denote $(k_{\pi^{-1}(1)}, \cdots, k_{\pi^{-1}(t)})^T$ by $\mbf{k}_{\pi}$, and define a state $\ket{\Psi_{\pi}} \in \mc{H}^{\otimes 2t}$ by
\begin{align}
\ket{\Psi_{\pi}} &:= I \otimes V(\pi) \ket{\Phi}\\
&=\frac{1}{d^{t/2}} \sum_{\mbf{k} \in \mc{N}} \ket{ \mbf{k}, {\mbf{k}^*_{\pi}}}\\
&=\frac{1}{d^{t/2}} \sum_{\blds{\alpha} \in \mc{N}} \ket{ \blds{\alpha}, {\blds{\alpha}^*_{\pi}}},
\end{align}
where $V(\pi)$ is a unitary representation of $\pi$, $ \ket{\Phi}$ is the maximally entangled state between the first $\mc{H}^{\otimes t}$ and the second $\mc{H}^{\otimes t}$,
$\ket{ \mbf{k}, {\mbf{k}^*_{\pi}}}=\ket{ \mbf{k}} \otimes (\ket{ \mbf{k}_{\pi}})^*$ and 
$\ket{ \blds{\alpha}, \blds{\alpha}^*_{\pi}}=\ket{ \blds{\alpha}} \otimes (\ket{ \blds{\alpha}_{\pi}})^*$.
Note that $\ket{\Psi_{\pi}}$ and $\ket{\Psi_{\sigma}}$ are not necessarily orthogonal depending on the permutation element. 
We denote $\ketbra{\Psi_{\pi}}{\Psi_{\pi}}$ simply by $\Psi_{\pi}$.

%Indeed, an inner product between them is given by
%\begin{align}
%\braket{\Psi_{\pi}}{\Psi_{\sigma}} &= \frac{1}{d^t} \sum_{\mbf{k}\in \mc{N}} \braket{\mbf{k}_{\sigma}}{\mbf{k}_{\pi}}\\
%&=
%\begin{cases}
%1 & \text{if $\pi = \sigma$}\\
%\frac{1}{d^t}
%\text{(\# of $\mbf{k} \in \mc{N}$ s.t. $\mbf{k}_{\sigma}= \mbf{k}_{\pi}$)} & \text{otherwise.}
%\end{cases}
%\end{align}

We also introduce three subspaces in $\mc{H}^{\otimes 2t}$,  
\begin{align}
&\mc{H}_{E} = {\rm span}\{ \ket{ \mbf{k}, \mbf{k}^*_{\pi}} :  \mbf{k} \in \mc{N}, \pi \in S_t\}, \\
&\mc{H}_{F} = {\rm span}\{ \ket{\blds{\alpha}, \blds{\alpha}^*_{\pi}} :  \blds{\alpha} \in \mc{N}, \pi \in S_t\}, \\
&\mc{H}_0 = {\rm span}\{ \ket{\Psi_{\pi}} :  \pi \in S_t \}.
\end{align}
Obviously, $\mc{H}_E \supsetneq \mc{H}_{0}$ and $\mc{H}_F \supsetneq \mc{H}_{0}$.
The projectors onto the subspaces $\mc{H}_E$, $\mc{H}_F$ and $\mc{H}_0$ are denoted by $P_E$, $P_F$ and $P_0$, respectively. 
We further introduce an equivalent relation $\sim_{\blds{k}}$ ($\blds{k} \in \mc{N}$) in $S_t$ such that $\pi \sim_{\blds{k}} \sigma$ if and only if $\mbf{k}_{\pi} =  \mbf{k}_{\sigma}$. A set of representative elements in equivalence classes by $\sim_{\blds{k}}$ is denoted by $S_t^{\mbf{k}}$. Using this notation, the projectors $P_E$ and $P_F$ are explicitly given by
\begin{align}
P_E &= \sum_{\mbf{k} \in \mc{N}} \sum_{\pi \in S^{\mbf{k}}_t} \ketbra{ \mbf{k},  \mbf{k}^*_{\pi}}{ \mbf{k},  \mbf{k}^*_{\pi}}, \\
P_F &= \sum_{\blds{\alpha} \in \mc{N}} \sum_{\pi \in S^{\blds{\alpha}}_t} \ketbra{ \blds{\alpha},  \blds{\alpha}^*_{\pi}}{ \blds{\alpha}, \blds{\alpha}^*_{\pi}} \label{PFrnp}
\end{align}
We have the following lemmas for these projectors.

\begin{Lemma}[Ref.~\cite{BHH2012,NKM2014}] \label{Lemma:A}
{\it For Haar random unitaries $U$, random diagonal-unitaries $D^E$ in the basis of $E$, and those $D^F$ in the basis of $F$, the following hold
\begin{align}
&\mathbb{E}_{U \sim {\sf H}}[U^{\otimes t,t}]  = P_0,\\
&\mathbb{E}_{D^E \sim {\sf D}_E}[(D^E)^{\otimes t,t} ] = P_E,\\
&\mathbb{E}_{D^F \sim {\sf D}_F}[(D^F)^{\otimes t,t} ] = P_F.
\end{align}
}
\end{Lemma}

\begin{Lemma}[Ref.~\cite{BHH2012}] \label{Lemma:B}
{\it It holds that $|\!| P_0  - \sum_{\pi \in S_t} \Psi_{\pi} |\!|_{\infty} \leq \frac{t^2}{d}$.
}
\end{Lemma}

\subsection{Proof of Theorem~\ref{Thm:main}} \label{Sec:Pmain}
We now prove Theorem~\ref{Thm:main}.
Due to the independence of random diagonal unitaries $D^E$, $D'^E$ and $D^F$ and Lemma~\ref{Lemma:A}, we have
\begin{equation}
|\! | \mathbb{E}_{D^E, D'^E\sim {\sf D}_E, D^F \sim {\sf D}_F}[(D^E D^F D'^E)^{\otimes t,t}] - \mathbb{E}_{U \sim {\sf H}}[U^{\otimes t,t}] |\! |_{\infty}=|\! | P_E P_F P_E -P_0 |\! |_{\infty},
\end{equation}
which is bounded from above as follows:
\begin{align}
|\!| P_E P_F P_E - P_0 |\!|_{\infty} 
&\leq
|\!| P_E P_FP_E - \sum_{\pi \in S_t} \Psi_{\pi} |\!|_{\infty}
+
|\!|P_0  - \sum_{\pi \in S_t} \Psi_{\pi} |\!|_{\infty}\\
%&=
%|\!| P_E (P_F  - \sum_{\pi \in S_t} \Psi_{\pi}) |\!|_{\infty}
%+
%|\!| P_0  - \sum_{\pi \in S_t} \Psi_{\pi} |\!|_{\infty}\\
&\leq
|\!| P_E (P_F  - \sum_{\pi \in S_t} \Psi_{\pi}) P_E|\!|_{\infty}
+\frac{t^2}{d}, \label{Eq:gar}
\end{align}
where we have used the triangular inequality in the first line,
the fact that $\ket{\Psi_{\pi}} \in \mc{H}_0 \subset \mc{H}_E$ and Lemma~\ref{Lemma:B} in the second line.
Using the fact that the operator norm for Hermitian operators is bounded from above by the row norm, defined by $\max_j \sum_i |A_{ij}|$ for an Hermitian operator $A$, we have 
\begin{align}
|\! | \mathbb{E}_{D^E, D'^E\sim {\sf D}_E, D^F \sim {\sf D}_F}[(D^E D^F D'^E)^{\otimes t,t}] - \mathbb{E}_{U \sim {\sf H}}[U^{\otimes t,t}] |\! |_{\infty}&\leq  C,
\end{align}
where 
\begin{equation}
C=
\max_{\mbf{l} \in \mc{N}}
\max_{\sigma \in S_t^{\mbf{l}}}
\sum_{\mbf{k} \in \mc{N}}
\sum_{\chi \in S_t^{\mbf{k}}}
\bigl| \bra{ \mbf{l}, \mbf{l}^*_{\sigma}}
P_F  - \sum_{\pi \in S_t} \Psi_{\pi}
\ket{ \mbf{k}, \mbf{k}^*_{\chi}} \bigr| + \frac{t^2}{d}. \label{Eq:defC}
\end{equation}
Note that it suffices to consider only vectors in $\mc{H}_E^{\otimes t,t}$ when we compute the first term of Eq.~\eqref{Eq:gar}, which is because the operator is sandwiched by the projector $P_E$. In the following, we evaluate $C$.

Substituting $\ket{\Psi_{\pi}}=\frac{1}{\sqrt{d^t}} \sum_{\mbf{m} \in \mc{N}} \ket{ \mbf{m}, \mbf{m}^*_{\pi}}$, the second term is given by
\begin{align}
\bra{ \mbf{l}, \mbf{l}^*_{\sigma}} \sum_{\pi \in S_t} \Psi_{\pi} \ket{ \mbf{k}, \mbf{k}^*_{\chi}}
&=
\frac{1}{d^{t}}
\sum_{\pi \in S_t} 
\delta_{\mbf{l}_{\pi}, \mbf{l}_{\sigma}} 
\delta_{\mbf{k}_{\pi}, \mbf{k}_{\chi}}.
\end{align}
On the other hand, using an explicit form of $P_F$ given in Eq.~\eqref{PFrnp}, the first term can be expanded to be
\begin{align}
\bra{\mbf{l},\mbf{l}^*_{\sigma}} P_F \ket{\mbf{k},\mbf{k}^*_{\chi}}
&=
\sum_{\blds{\alpha} \in \mc{N}} \sum_{\pi \in S_t^{\blds{\alpha}}}
\braket{\mbf{l}}{\blds{\alpha}}
\braket{\blds{\alpha}}{\mbf{k}}
\braket{\mbf{k}_{\pi^{-1} \circ \chi}}{\blds{\alpha}}
\braket{\blds{\alpha}}{\mbf{l}_{\pi^{-1} \circ \sigma}}.
\end{align}
Since a pair of the bases $(E,F)$ is a Fourier-type pair, it satisfies for any $l,k,\alpha \in [0,d-1]$ that
$\braket{l}{\alpha}\braket{k}{\alpha}
= \braket{l + k}{\alpha}/d^{1/2}$, where $l+k \in [0,d-1]$ as $[0,d-1]$ is an additive group with respect to $+$ .
Denoting $( l_1 + k_1, \cdots, l_t + k_t)^T$ by $\mbf{l} + \mbf{k}$, we have
\begin{align}
\bra{\mbf{l},\mbf{l}_{\sigma}^*} P_F \ket{\mbf{k},\mbf{k}_{\chi}^*}
&=
\frac{1}{d^{t}}
\sum_{\blds{\alpha} \in \mc{N}} \sum_{\pi \in S_t^{\blds{\alpha}}}
\braket{\mbf{l} + \mbf{k}_{\pi^{-1} \circ\chi}}{\blds{\alpha}}
\braket{\blds{\alpha}}{\mbf{k}+\mbf{l}_{\pi^{-1} \circ \sigma}}\\
&=
\frac{1}{d^{t}}
\sum_{\blds{\alpha}\in \mc{N}} \biggl(\sum_{\pi \in S_t} - \sum_{\pi \in S_t \setminus S_t^{\blds{\alpha}}} \biggr)
\braket{\mbf{l} + \mbf{k}_{\pi^{-1} \circ\chi}}{\blds{\alpha}}
\braket{\blds{\alpha}}{\mbf{k}+\mbf{l}_{\pi^{-1} \circ \sigma} }\\
&=
\frac{1}{d^{t}}
\biggl(
\sum_{\pi \in S_t}
\delta_{\mbf{l} + \mbf{k}_{\pi^{-1} \circ\chi}, \mbf{k}+\mbf{l}_{\pi^{-1} \circ \sigma} }
-
M_{\mbf{l},\mbf{k}}\biggr),
\end{align}
where $M_{\mbf{l},\mbf{k}}=\sum_{\blds{\alpha} \in \mc{N}} \sum_{\pi \in S_t \setminus S_t^{\blds{\alpha}}} \braket{\mbf{l} + \mbf{k}_{\pi^{-1} \circ\chi}}{\blds{\alpha}}
\braket{\blds{\alpha}}{\mbf{k}+\mbf{l}_{\pi^{-1} \circ \sigma} }$, 
and we used $\sum_{\blds{\alpha} \in \mc{N}} \ketbra{\blds{\alpha}}{\blds{\alpha}} =I_{\mc{H}^{\otimes t}}$.
Hence, using the triangular inequality, we obtain
\begin{align}
|\bra{\mbf{l},\mbf{l}_{\sigma}^*}
P_F  - \sum_{\pi \in S_t} \Psi_{\pi}
\ket{\mbf{k},\mbf{k}_{\chi}^*}| 
&=
\frac{1}{d^t}
\biggl|
\sum_{\pi \in S_t}
\bigl(\delta_{\mbf{l} + \mbf{k}_{\pi^{-1} \circ\chi},\mbf{k}+\mbf{l}_{\pi^{-1} \circ \sigma}}
-
\delta_{\mbf{l}_{\pi}, \mbf{l}_{\sigma}} 
\delta_{\mbf{k}_{\pi}, \mbf{k}_{\chi}}\bigr)
-
M_{\mbf{l}, \mbf{k}}
\biggr| \\
&\leq
\frac{1}{d^t}
\biggl|
\sum_{\pi \in S_t}
\bigl(\delta_{\mbf{l} + \mbf{k}_{\pi^{-1} \circ\chi},\mbf{k}+\mbf{l}_{\pi^{-1} \circ \sigma}}
-
\delta_{\mbf{l}_{\pi}, \mbf{l}_{\sigma}} 
\delta_{\mbf{k}_{\pi}, \mbf{k}_{\chi}} \bigr) \biggr|
+
\frac{1}{d^t}
\bigl| M_{\mbf{l},\mbf{k}} \bigr|.
\end{align}

An upper bound of $| M_{\mbf{l},\mbf{k}} \bigr|$ can be obtained from the fact that the bases $E$ and $F$ are mutually unbiased, leading to
\begin{equation}
\bigl| M_{\mbf{l},\mbf{k}} \bigr| \leq \frac{1}{d^t} \sum_{\blds{\alpha} \in \mc{N}}| S_t \setminus S_t^{\blds{\alpha}}|.
\end{equation}
As $| S_t \setminus S_t^{\blds{\alpha}}|$ depends only on how many different elements $\blds{\alpha}$ contains, the number of which we denote by $k$, and the number of every different element $\alpha_i$ in $\blds{\alpha}$, denoted by $s_i$,
we replace the summation with that over $k$ and obtain
\begin{equation}
\sum_{\blds{\alpha} \in \mc{N}}
\bigl| S_t \setminus S_t^{\blds{\alpha}} \bigr|
=
\sum_{k=1}^{t}\binom{d}{k} g^{(k)}(t), 
\end{equation}
where the binomial coefficient counts the number of possible choices of $k$ different numbers from $[0,d-1]$, and $g^{(k)}(t)$ is the function that depends only on $k$ and $t$ given by
\begin{equation}
g^{(k)}(t) = \sum_{(s_1,\cdots,s_k)} \frac{t!}{s_1! \cdots s_k!} \bigl( t! -\frac{t!}{s_1! \cdots s_k!} \bigr). 
\end{equation}
Here, the summation is taken over all possible $(s_1,\cdots,s_k)$ such that $\forall i \in [1,k]$ $s_i \in [1, t]$ and $\sum_{i=1}^k s_i =t$. 
For a fixed $k$, the number of such combinations is simply given by $\binom{t-1}{k-1}$.
For $k=t$, $s_i =1$ for all $i\in [1,k]$ and thus $g^{(t)}(t)=0$. 
%This can be also observed from a fact that $| S_t \setminus S_t^{\blds{\alpha}}|=0$ for $\blds{\alpha} = (\alpha_1, \cdots, \alpha_t)$ that satisfies $\alpha_p \neq \alpha_q (p \neq q)$.
For the remaining terms $g^{(k)}(t)$ ($k \in [1,t-1]$), we use an upper bound given by
\begin{equation}
g^{(k)}(t) \leq \binom{t-1}{k-1} \frac{t!^2}{4},
\end{equation}
which is optimal when $k=t-1$. Substituting these, we obtain
\begin{align}
\sum_{\blds{\alpha} \in \mc{N}}
\bigl| S_t \setminus S_t^{\blds{\alpha}} \bigr|
&\leq
\frac{t!^2}{4} \sum_{k=1}^{t-1}\binom{d}{k}\binom{t-1}{k-1}\\
&=\frac{t!^2}{4} \biggl( \binom{d-1+t}{t} - \binom{d}{t} \biggr),
\end{align}
where the last line is obtained due to the Vandermonde's identity.
Since $d = \Omega(t^2)$, an upper bound is obtained such as
\begin{align}
\sum_{\blds{\alpha} \in \mc{N}}
\bigl| S_t \setminus S_t^{\blds{\alpha}} \bigr|
&\leq t^2 t! d^{t-1} + O(t^4 t! d^{t-2}),
\end{align}
which leads to 
\begin{align}
|\bra{\mbf{l},\mbf{l}^*_{\sigma}}
P_F  - \sum_{\pi \in S_t} \Psi_{\pi}
\ket{\mbf{k},\mbf{k}^*_{\chi}}| 
&\leq
\frac{1}{d^t}
\biggl|
\sum_{\pi \in S_t}
\bigl(\delta_{\mbf{l} + \mbf{k}_{\pi^{-1} \circ\chi},\mbf{k}+\mbf{l}_{\pi^{-1} \circ \sigma}}
-
\delta_{\mbf{l}_{\pi}, \mbf{l}_{\sigma}} 
\delta_{\mbf{k}_{\pi}, \mbf{k}_{\chi}} \bigr) \biggr|
+
\frac{t^2 t!}{d^{t+1}}+O\biggl(\frac{t^4 t!}{d^{t+2}} \biggr).
\end{align}
Substituting this into $C$, the following upper bound can be obtained:
\begin{align}
C &\leq 
\frac{t^2(t!^2+1)}{d}+
\frac{1}{d^t}
\max_{\mbf{l} \in \mc{N}}
\max_{\sigma \in S_t^{\mbf{l}}}
\sum_{\mbf{k} \in \mc{N}}
\sum_{\chi \in S_t^{\mbf{k}}}
\biggl|
\sum_{\pi \in S_t} 
\bigl(\delta_{\mbf{l} + \mbf{k}_{\pi^{-1} \circ\chi},\mbf{k}+\mbf{l}_{\pi^{-1} \circ \sigma}}
-
\delta_{\mbf{l}_{\pi}, \mbf{l}_{\sigma}} 
\delta_{\mbf{k}_{\pi}, \mbf{k}_{\chi}} \bigr) 
\biggr| 
+ O\biggl(\frac{t^4  t!^2}{d^{2}} \biggr)\\
&=
\frac{t^2(t!^2+1)}{d}+
\frac{1}{d^t}
\max_{\mbf{l} \in \mc{N}}
\max_{\sigma \in S_t^{\mbf{l}}}
\sum_{\mbf{k} \in \mc{N}}
\sum_{\chi \in S_t^{\mbf{k}}}
\sum_{\pi \in S_t} 
\bigl(\delta_{\mbf{l} + \mbf{k}_{\pi^{-1} \circ\chi},\mbf{k}+\mbf{l}_{\pi^{-1} \circ \sigma}}
-
\delta_{\mbf{l}_{\pi}, \mbf{l}_{\sigma}} 
\delta_{\mbf{k}_{\pi}, \mbf{k}_{\chi}} \bigr)  
+ O\biggl(\frac{t^4  t!^2}{d^{2}} \biggr)\\
&\leq 
\frac{t^2(t!^2+1)}{d}+
\frac{1}{d^t}
\max_{\mbf{l} \in \mc{N}}
\max_{\sigma \in S_t^{\mbf{l}}}
\sum_{\pi \in S_t} 
\sum_{\mbf{k} \in \mc{N}}
\sum_{\chi (\neq \pi) \in S_t^{\mbf{k}}}
\delta_{\mbf{l} + \mbf{k}_{\pi^{-1} \circ\chi},\mbf{k}+\mbf{l}_{\pi^{-1} \circ \sigma}}
+ O\biggl(\frac{t^4  t!^2}{d^{2}} \biggr) \\
&\leq 
\frac{t^2(t!^2+1)}{d}+
\frac{1}{d^t}
\max_{\mbf{l} \in \mc{N}}
\max_{\sigma \in S_t^{\mbf{l}}}
\sum_{\pi \in S_t} 
\sum_{\chi (\neq \pi) \in S_t}
\sum_{\mbf{k} \in \mc{N}}
\delta_{\mbf{l} + \mbf{k}_{\pi^{-1} \circ\chi},\mbf{k}+\mbf{l}_{\pi^{-1} \circ \sigma}}
+ O\biggl(\frac{t^4  t!^2}{d^{2}} \biggr), 
\label{Eq:done}
\end{align}
where the second line is due to a fact that the term in the modulus is non-negative because, when the second term is one, the first term is also one, the 
third line is obtained by using a fact that the first and the second term cancel each other when $\chi=\pi$ and by dropping negative terms when $\chi \neq \pi$, and the last line is due to $S_t^{\mbf{k}} \subset S_t$.
For the delta function $\delta_{\mbf{l} + \mbf{k}_{\pi^{-1} \circ\chi},\mbf{k}+\mbf{l}_{\pi^{-1} \circ \sigma}}$, we have 
\begin{equation}
\delta_{\mbf{l} + \mbf{k}_{\pi^{-1} \circ\chi},\mbf{k}+\mbf{l}_{\pi^{-1} \circ \sigma}}=1 
\Longleftrightarrow \forall s \in [1, t], \ \ \ \  l_s + k_{\chi^{-1} \circ \pi(s)}= k_s + l_{\sigma^{-1} \circ \pi(s)}
\end{equation}
When $\chi \neq \pi$, there exists at least one pair $(s,s')$ ($s \neq s' \in [1,t]$) such that $\pi(s) = \chi(s')$. Hence, $k_{s'} = k_s + l_{\sigma^{-1} \circ \pi(s)} - l_s$ should be at least satisfied for the delta function to be non-zero. Thus, the number of $\mbf{k}$, for which the delta function is non-zero, is at most $d^{t-1}$.
Based on this observation, we obtain
\begin{equation}
\max_{\mbf{l} \in \mc{N}}
\max_{\sigma \in S_t^{\mbf{l}}}
\sum_{\pi \in S_t} 
\sum_{\chi (\neq \pi) \in S_t}
\sum_{\mbf{k} \in \mc{N}}
\delta_{\mbf{l} + \mbf{k}_{\pi^{-1} \circ\chi},\mbf{k}+\mbf{l}_{\pi^{-1} \circ \sigma}}
\leq
t!^2 d^{t-1}.
\end{equation}
Substituting this into Eq.~\eqref{Eq:done}, we obtain an upper bound of $C$, leading to
\begin{equation}
|\! | \mathbb{E}_{D^E, D'^E\sim {\sf D}_E, D^F \sim {\sf D}_F}[(D^E D^F D'^E)^{\otimes t,t}] - \mathbb{E}_{U \sim {\sf H}}[U^{\otimes t,t}] |\! |_{\infty} \leq \frac{(1+t^2)t!^2+t^2}{d} + O\biggl(\frac{t^4  t!^2}{d^{2}} \biggr).
\end{equation}
This concludes the proof. $\hfill \blacksquare$

\subsection{Proof of Lemma~\ref{Lemma:PairsLP}} \label{Sec.PPairs}

We first provide a key lemma to prove Lemma~\ref{Lemma:PairsLP}. 
The lemma is seen as a constrained problem in extremal algebraic theory~\cite{AAK2003,AAK2003-2}.
The proof is given in Appendix~\ref{Sec:ProofCounting}.

\begin{Lemma} \label{Lemma:counting}
\emph{
Let $O$ be an orthogonal matrix acting on the Euclidean space $\mbb{R}^t$, which contains the set of apexes of a hypercube, $\{ 0,1\}^t$. 
If there exists a set $S \subset \{0,1 \}^t$ such that $O S \subset \{0,1\}^t$ and $|S| > 2^{t-1}$, then $O$ is a permutation matrix.
}
\end{Lemma}

Now, we prove Lemma~\ref{Lemma:PairsLP}, i.e. $\Lambda_2 =|L_2| \leq 2^{2t^2 +(t-1) N}$.
Here, $L_2$ is the set of pairs $(K, K')$, where $K$ is a $2$-local but not a row permutation of $K'$.

\begin{Proof}[Lemma~\ref{Lemma:PairsLP}]
Throughout the proof, we denote the column vectors of $K$ and $K'$ by $\vec{k}_i$ and $\vec{k}'_i$, respectively, for $i \in [1,N]$.
The $2$-local permutation condition is equivalent to the following:
\begin{equation}
\forall i,j \in [1,N], \ \  \vec{k}_i \cdot \vec{k}_j=\vec{k}'_i \cdot \vec{k}'_j, \label{mkq5y}
\end{equation}
where $\cdot$ is the usual Euclidean inner product.
This is because the conditions for $i=j$ imply that the number of $1$'s in $\vec{k}_i$ and that in $\vec{k}'_i$ should be the same, and those for $i\neq j$ imply that the number of $11$ in $K_{\{i,j\}}$ is equal to that in $K'_{\{i,j\}}$. 
These conditions together correspond to the necessary and sufficient conditions for the pair $(K, K')$ to be $2$-local permutations.
Moreover, Eq.~\eqref{mkq5y} implies that the Gram matrix of a set $\{\vec{k}_i: i \in [1,N]\}$ of column vectors is the same as that of $\{\vec{k}'_i: i \in [1,N]\}$.
Hence, ${\rm span}\{\vec{k}_i: i \in [1,N]\}$ has the same dimension as ${\rm span}\{\vec{k}'_i: i \in [1,N]\}$. 
It also follows that there exists a partial isometry $O$ that satisfies $O \vec{k}_i= \vec{k}'_i$ for any $i \in [1,N]$, i.e. $O K = K'$. 
If the partial isometry is restricted to its support, it is an orthogonal matrix as the elements of the vectors are in $\{0,1\}$, and it is not a permutation operator due to the assumption that $K$ is not a row permutation of $K'$.

We will now construct a set $\mc{O}$ of orthogonal matrices on $\mbb{R}^t$ that satisfies 
\begin{equation}
\forall (K,K') \in L_2,   \exists O \in \mc{O} \text{\  such  that\ } O K=K'. \label{Eq:zmdosf}
\end{equation} 
This can be done as follows.
Let $s:= 2^{2t}$ and $[0,s-1]^{\leq t}$ be the set of $s$-ary strings of length $t$ or smaller. 
We describe a procedure of defining a set $S_2 \subset [0,s-1]^{\leq t}$ and orthogonal matrices $O_{\bf{b}}$ for $\mbf{b} \in S_2$, such that $S_2$ is a prefix code and that $\mc{O}:= \{O_{\mbf{b}}|\mbf{b} \in S_2\}$ satisfies Eq.~\eqref{Eq:zmdosf}.
Our construction starts with $S_2=\emptyset$ and is recursive in terms of the rank $\kappa$ of the partial isometry obtained from $(K,K')$.
We repeat the subroutine described below from $\kappa = t$ to $\kappa=1$ by decreasing $\kappa$ one by one.
In the subroutine, we first choose $(K, K') \in L_2$ that defines a partial isometry with rank $\kappa$. We pick up an arbitrary set of independent column vectors $\{\vec{k}_{i_m}\}_{m=1}^{\kappa}$ in $K$ and those $\{\vec{k}'_{i_m}\}_{m=1}^{\kappa}$ in $K'$. These vectors can be converted to an $s$-ary string $\mbf{b}= (2^t k_{i_1}+k'_{i_1}, 2^tk_{i_2}+k'_{i_2}, \cdots , 2^tk_{i_{\kappa}}+k'_{i_{\kappa}})$ of length $\kappa$ by regarding each vector as a binary number with length $t$. 
If $\mbf{b}$ is a prefix of a string $\mbf{b}' \in S_2$, then 
the orthogonal matrix $O_{\mbf{b}'}$ satisfies $O_{\mbf{b}'}K = K'$ because, on the support of the partial isometry obtained from $(K,K')$, the action of $O_{\mbf{b}'}$ is the same as that of the isometry by construction. Otherwise, we append $\mbf{b}$ to $S_2$, and define an orthogonal matrix $O_{\mbf{b}}$ as an arbitrary extension of the partial isometry. The subroutine is run for all $(K, K') \in L_2$ with a partial isometry of rank $\kappa$.
Eventually, we obtain a set $\mc{O}$ of orthogonal matrices on $\mbb{R}^t$. 
Importantly, it does not contain a permutation matrix and, by construction, $|\mc{O}|=|S_2| \leq 2^{2 t^2}$. 

Introducing a set $L_2(O)$ by $\{ (K, O K) : K, OK \in \{0,1\}^{t N}\}$ for a given orthogonal matrix $O \in \mbb{R}^t$,
we have  $L_2 \subset \bigcup_{O \in \mc{O}} L_2(O)$, leading to 
\begin{align}
\Lambda_2 &\leq \sum_{O \in \mc{O}} | L_2(O)|\\
&\leq |\mc{O}| \max_{O \in \mc{O}} | L_2(O)|\\
&\leq 2^{2t^2} \max_{O \in \mc{O}} | L_2(O)|.
\end{align}
Since the condition $OK \in \{0,1 \}^{tN}$ consists of an identical and independent condition on each column of $K$, $|L_2(O)|$ for $O \in \mc{O}$ is bounded from above by 
\begin{equation}
|L_2(O)| \leq 
\biggl(  \max_{O \in \mc{O}} \bigl|\{\vec{k} \in \{0,1\}^t:  O\vec{k} \in \{0,1\}^t \} \bigr|\biggr)^N.
\end{equation}
As $O \in \mc{O}$ is on $\mathbb{R}^t$ and is not a permutation matrix, from the contraposition of Lemma~\ref{Lemma:counting}, we obtain
\begin{equation}
\max_{O \in \mc{O} } \bigl|\{\vec{k} \in \{0,1\}^t:  O\vec{k} \in \{0,1\}^t \bigr|
\leq 2^{t-1}.
\end{equation}
Thus, we have
$\Lambda_2 \leq 2^{2t^2 + (t-1) N}$, and conclude the proof. $\hfill \blacksquare$
\end{Proof}

Finally, we note that the upper bound given in Lemma~\ref{Lemma:PairsLP} is unlikely to be tight in terms of $t$ because an upper bound $|\mc{O}|$ given by $2^{2t^2}$ in the proof is far from optimal. This is observed from the fact that $|\mc{O}| = |S_2|$ but $S_2$ does not contain all strings with length $t$.
More concretely, we provide instances for a small $t$.
From the result in Ref.~\cite{NKM2014}, we know that, for any pair $(K,K')$,
$K$ is a row permutation of $K'$ if and only if $K$ is a $(\lfloor \log_2 t \rfloor +1)$-local permutation of $K'$. 
Hence, the smallest $t$ making the $2$-local permutation check problem non-trivial is $4$. In this case, we can show that, if $K$ is a $2$-local but  not a row permutation of $K'$,
the four rows of $K$ and those of $K'$ can be rearranged independently, resulting in $K_{\pi}$ and $K'_{\sigma}$ respectively ($\pi, \sigma \in S_4$), 
such that a pair of the $i$th column of $K_{\pi}$ and that of $K'_{\sigma}$ are in the set $C_0 \cup C_1$ ($\forall i \in [1,N]$), where
\begin{align}
C_0 &= \bigl\{\bigl( (0,0,0,0)^T, (0,0,0,0)^T \bigr), \bigl( (1,1,1,1)^T, (1,1,1,1)^T \bigr),\bigl( (0,0,1,1)^T, (0,0,1,1)^T \bigr),\notag \\
&\ \ \ \ \ \ \ \ \bigl( (1,1,0,0)^T, (1,1,0,0)^T \bigr),\bigl( (1,0,1,0)^T, (1,0,1,0)^T \bigr),\bigl( (0,1,0,1)^T, (0,1,0,1)^T \bigr) \bigr\},\\
C_1 &= \bigl\{\bigl( (0,1,1,0)^T, (1,0,0,1)^T \bigr),\bigl( (1,0,0,1)^T, (0,1,1,0)^T \bigr) \bigr\}.
\end{align}
Taking the number of choices of $\pi$ and $\sigma$ into account, we have
\begin{equation}
\Lambda_2 < t!^2  \bigl( |C_0|+|C_1| \bigr)^N = t!^2 8^{N},
\end{equation} 
which corresponds to $t!^2 2^{(t-1)N}$ for $t =4$.
For this reason, we conjecture that the optimal bound should be given by $f(t) 2^{(t-1)N}$ where $f(t) = O({\rm poly}(t!))$, which we have analytically confirmed for $t \leq 7$.
If this conjecture is true, Theorem~\ref{Thm:2} works for $t = o(N/\log N)$ instead of $t = o(N^{1/2})$.

\subsection{Proof of Corollary~\ref{Cor:3}} \label{Sec:HamDes}

We prove Corollary~\ref{Cor:3} that, $\forall T \geq (2t+1 + \frac{2}{N} \log_2 1/\epsilon)\pi $,
a random unitary $U_{XZ}(T) = \mc{T}\exp[-i \int_0^T ds H_{XZ}(s)]$ generated by $H_{XZ}(T) \in_{\rm R} \mathfrak{H}_{XZ}^{(t)}$ at time $T$ is an $\epsilon$-approximate unitary $t$-design, where $\mathfrak{H}_{XZ}^{(t)}$ is the set of Hamiltonians in the form of Eq.~\eqref{Eq:designHam}.
%
%We prove Corollary. To remind the statement, let $\mathfrak{H}_{XZ}^{(t)}$ be the set of $2$-local time-dependent Hamiltonians in the form of
%\begin{equation}
%H_{XZ} (T) = \begin{cases}
%H_Z^{(m)} & \text{if\ \  $ 2m \pi \leq T < (2m+1) \pi$},\\
%H_X^{(m)} & \text{if\ \  $  (2m+1) \pi \leq  T < 2(m+1) \pi$},
%\end{cases}
%\end{equation}
%over $H_W^{(m)} \in  \mathfrak{H}_W^{(t)}$ ($W=X,Z$ and $m =0,1,\dots$), which are defined in Eqs.~\eqref{HamX} and~\eqref{HamZ}.
%Then, we show that $\forall T \geq (2t+1 + \frac{2}{N} \log 1/\epsilon)\pi $,
%a random unitary $U_{XZ}(T) = \mc{T}\exp[-i \int_0^T ds H_{XZ}(s)]$ generated by $H_{XZ}(T) \in_{\rm R} \mathfrak{H}_{XZ}^{(t)}$ at time $T$
%is an $\epsilon$-approximate unitary $t$-design.

\begin{Proof}[Corollary~\ref{Cor:3}]
In the proof, we denote $e^{-i \tau H_W^{(m)}}$ by $U_W^{(m)}(\tau)$ ($W=X,Z$). 
As both Hamiltonians are composed of commuting terms, they are simply given by
\begin{equation}
e^{-i  \tau H_X^{(m)}} = \prod_{k<k'} e^{i \tau J_{kk'}^{(m)} X_k \otimes X_{k'}} \prod_k e^{i \tau B_{k}^{(m)} X_k} \ \ \  {\rm and } \ \ \ 
e^{-i  \tau H_Z^{(m)}} = \prod_{k<k'} e^{i \tau \tilde{J}_{kk'}^{(m)} Z_k \otimes Z_{k'}} \prod_k e^{i \tau \tilde{B}_{k}^{(m)} Z_k}.
\end{equation}

We first consider a random unitary $U_{XZ}(T_{\ell})$ at time $T_{\ell}= (2 \ell+1) \pi$ ($\ell=1,2,\dots$). Using the above notation, it is given by
\begin{equation}
U_{XZ}(T_{\ell})
=
U_Z^{(\ell+1)}(\pi) \prod_{m=\ell}^{1} U_X^{(m)}(\pi)U_Z^{(m)}(\pi). \label{mq34:p}
\end{equation}
We take the average of $U_{XZ}(T_{\ell})^{\otimes t, t}$ over $H_{XZ} \in_{\rm R} \mathfrak{H}_{XZ}^{(t)}$, which is equivalent to take the average over all parameters $B_k^{(m)}, \tilde{B}_{k'}^{(m)} \in_{\rm R} \mc{P}_t(B)$ and $J_{kk'}^{(m)}, \tilde{J}_{kk'}^{(m)} \in_{\rm R} \mc{P}_t(J)$.
Here, the parameter set $\mc{P}_t(c)$ is given by Eq.~\eqref{Eq:parameter} such as
\begin{align}
\mc{P}_t(c) &= \biggl\{ \frac{m}{2 (\lfloor t/2 \rfloor + 1)}: m \in \bigl[-c, c \bigr] \biggr\},
\end{align}
and $(B,J)=(\lfloor t/2 \rfloor+1/2, \lfloor t/2 \rfloor/2)$. Since it holds that
\begin{widetext}
\begin{multline}
e^{i \pi J_{kk'}^{(m)} Z_k \otimes Z_{k'}} \ 
e^{i \pi B_{k}^{(m)} Z_k} \otimes e^{i \pi B_{k'}^{(m)} Z_{k'}} \\
=
e^{\pi i (J_{k k'}^{(m)} + B_k^{(m)}+ B_{k'}^{(m)})}
\bigl( {\rm diag}_Z\{1, e^{-2\pi i (J_{k k'}^{(m)}+ B_{k'}^{(m)})}\}
\otimes
{\rm diag}_Z\{1, e^{-2\pi i (J_{k k'}^{(m)}+ B_{k}^{(m)})}\} \bigr)
{\rm diag}_Z \{1,1,1,e^{4\pi i J_{k k'}^{(m)}} \}, \label{Eq:ha8sv9d}
\end{multline}
\end{widetext}
if $B_{k}^{(m)}, B_{k'}^{(m)} \in_{\rm R} \mc{B}_t$ and $J_{kk'}^{(m)} \in_{\rm R} \mc{J}_t$, where
\begin{align}
\mc{B}_t &= \biggl\{ \frac{m}{2 (\lfloor t/2 \rfloor+1)} : m \in [0, 2 \lfloor t/2 \rfloor+1 ]  \biggr\}\\
\mc{J}_t &= \biggl\{ \frac{m}{2 (\lfloor t/2 \rfloor+1)} : m \in [0, \lfloor t/2 \rfloor ]  \biggr\},
\end{align}
then the probability distribution of $(-2 \pi (J_{kk'}^{(m)}+B_{k'}), -2\pi (J_{kk'}^{(m)}+B_k^{(m)}), 4 \pi J_{kk'}^{(m)})$ is identical to that of
$(\varphi, \varphi', \theta)$ in Eq.~\eqref{Eq:RDCno;j4} with $a= 2(\lfloor t/2 \rfloor+1)$ and $b= \lfloor t/2 \rfloor+1$,
implying that $U_Z^{(m)}(T_{\ell})$ is equivalent to ${\rm RDC}_{\rm disc}(\mc{I}_2:2b,b)$ up to a global phase.
Noting that the global phase is cancelled in $U_Z^{(m)}(T_{\ell})^{\otimes t,t}$ and recalling that $\mathbb{E}[ {\rm RDC}_{\rm disc}(\mc{I}_2:a,b)^{\otimes t,t}] = \mathbb{E}[ {\rm RDC}(\mc{I}_2)^{\otimes t,t}]$
if $a \geq t+1$ and $b \geq \lfloor t/2 \rfloor +1$, we have
\begin{equation}
\mathbb{E}_{B_{k}^{(m)} \in_{\rm R} \mc{B}_t, J_{kk'}^{(m)} \in_{\rm R} \mc{J}_t}[ U_Z^{(m)}(T_{\ell})^{\otimes t,t} ] = \mathbb{E}[ {\rm RDC}(\mc{I}_2)^{\otimes t,t}]. \label{Eq:@@@}
\end{equation}
Using a product of two-qubit diagonal gates $V$ given by
\begin{equation}
V=\bigotimes_{k=1}^N {\rm diag}_Z^{(k)}\{ 1, e^{2 \pi i \Delta B} \}
\bigotimes_{k<k'} {\rm diag}_Z^{(kk')} \{ 1, 1,1, e^{-4 \pi i \Delta J} \},
\end{equation}
where the superscript of ${\rm diag}_Z$, such as $(k)$ and $(kk’)$, indicates the place of qubits the gate acts on, and $(\Delta B,\Delta J) = (\frac{\lfloor t/2 \rfloor+1/2}{2(\lfloor t/2 \rfloor+1)},\frac{\lfloor t/2 \rfloor}{4(\lfloor t/2 \rfloor+1)})$, 
we obtain
\begin{align}
\mathbb{E}_{B_{k}^{(m)} \in_{\rm R} \mc{P}_t(B), J_{kk'}^{(m)} \in_{\rm R} \mc{P}_t(J)}[ U_Z^{(m)}(T_{\ell})^{\otimes t,t} ]
&=\mathbb{E}_{B_{k}^{(m)} \in_{\rm R} \mc{B}_t, J_{kk'}^{(m)} \in_{\rm R} \mc{J}_t}[ U_Z^{(m)}(T_{\ell})^{\otimes t,t} ]V^{\otimes t,t} \\
&= \mathbb{E}[ {\rm RDC}(\mc{I}_2)^{\otimes t,t}] V^{\otimes t,t},
\end{align}
where we used Eq.~\eqref{Eq:@@@} in the last line.
Further, because ${\rm RDC}(\mc{I}_2)$ is composed of two-qubit diagonal gates with random phases uniformly drawn from $[0,2\pi)$, the average of ${\rm RDC}(\mc{I}_2)^{\otimes t,t}$ does not change even when additional diagonal two-qubit gates  are applied. Thus, we obtain
\begin{equation}
\mathbb{E}_{B_{k}^{(m)} \in_{\rm R} \mc{P}_t(B), J_{kk'}^{(m)} \in_{\rm R} \mc{P}_t(J)}[ U_Z^{(m)}(T_{\ell})^{\otimes t,t} ] = \mathbb{E}[ {\rm RDC}(\mc{I}_2)^{\otimes t,t}]. 
\end{equation}
As a similar relation holds for $X$ Hamiltonians, we conclude that
\begin{equation}
\mathbb{E}[U_{XZ}(T_{\ell})^{\otimes t,t}] = \mathbb{E}[\bigl( (RDC(\mc{I}_2) H_N )^{2\ell} RDC(\mc{I}_2) \bigr)^{\otimes t,t}],
\end{equation}
where $H_N$ is the Hadamard transformation on $N$ qubits, implying that $U_{XZ}(T_{\ell})$ is an $\epsilon$-approximate unitary $t$-design if $\ell \geq  t+ \frac{1}{N}\log_2 (1/\epsilon)$.

To complete the proof, consider the time $T$ satisfying $T_{\ell} < T < T_{\ell+1}$ where $\ell \geq  t+ \frac{1}{N}\log_2 (1/\epsilon)$. Because the time evolution operator from time $T_{\ell}$ to time $T$ is independent of the one before $T_{\ell}$, it follows from Lemma~\ref{Lemma:additionalU} that $U_{XZ}(T)$ is also an $\epsilon$-approximate unitary $t$-design. $\hfill \blacksquare$
\end{Proof}

\section{Conclusion and discussions} \label{Sec:Conclusion}

In this paper, we have presented new constructions of unitary $t$-designs and proposed design Hamiltonians as a general framework to investigate randomising operations in complex quantum many-body systems.
The new constructions are based on repetitions of random diagonal-unitaries in mutually unbiased bases. We have first shown that, if the bases are Fourier-type, approximate unitary $t$-designs can be achieved on one qudit after $O(t)$ repetitions. We have then constructed quantum circuits on $N$ qubits that achieve approximate unitary $t$-designs using  $O(tN^2)$ gates, which drastically improves the previous result~\cite{BHH2012} in terms of $t$.
The dependence on $t$ is essentially optimal amongst designs with finite supports.
The circuits were obtained by solving a special case of combinatorial problems, which we call the local permutation check problems, showing an interesting connection between combinatorics and efficient implementations of designs. Based on these results, we have provided a design Hamiltonian, which changes the interactions only a few times to generate designs. This result supports the natural design Hamiltonian conjecture and is also practically important as it simplifies the experimental implementations of unitary designs.

Our approach of studying unitary designs and randomising operations in physically natural systems opens a lot of interesting questions.
The following are a few questions concerning unitary designs:
\begin{enumerate} \vspace{-2mm}
\item In one-qudit systems, is it possible to implement unitary $t$-designs by repeating random diagonal-unitaries in \emph{any} non-trivial pairs of bases? If so, how many repetitions are sufficient for the implementations? \vspace{-2mm}
\item What is the best strategy of the local permutation check problems? \vspace{-2mm}
\item What is the most efficient implementation by quantum circuits that approximate random diagonal-unitaries in the Pauli-$Z$ basis? \vspace{-2mm}
\item What are the further applications of unitary $t$-designs for $t \geq 4$. \vspace{-2mm}
\end{enumerate}
Regarding the question 1, we have found that repeating random diagonal-unitaries in non-trivial pairs of bases achieves a unitary $1$-design if any vector in one basis is not orthogonal to any vector in the other basis. 
Although this non-orthogonality condition may not be necessary, we expect that, for arbitrary non-trivial pairs of bases satisfying the non-orthogonality condition, the process  eventually achieves unitary $t$-designs. 
The questions 2 and 3 are related each other due to Lemma~\ref{Lemma:LP}.
In this paper, we have considered only $2$-local permutation check problems.
However, if there exists a set $\mc{I} =\{ I\}$ such that $\Lambda(\mc{I}) = O(2^{(t-1)N})$ and $|I|=$ constant for all $I \in \mc{I}$, then we can implement approximate unitary $t$-designs using $O(t |\mc{I}|)$ quantum gates. Hence, finding a better strategy for the local permutation check problems immediately results in a faster implementation of unitary designs.
It is also desirable to directly search efficient quantum circuits approximating random diagonal-unitaries in the $Z$ basis because Lemma~\ref{Lemma:LP} may not be tight.
Finally, it is important to find applications of unitary $t$-designs for large $t$. A possible and promising direction is to further explore large deviation bounds for unitary designs as mentioned in Section~\ref{Sec:Nqubit}. 

We also list a few open questions about design Hamiltonians from the physical point of view:
\begin{enumerate}[I]
\item Prove or disprove the natural design Hamiltonian conjecture. \vspace{-2mm}
\item What are the exact relations between natural design Hamiltonians and various definitions of scrambling or OTO correlators? \vspace{-2mm}
\item If a design Hamiltonian is defined on a finite ensemble of local Hamiltonians,
how many Hamiltonians are needed? \vspace{-2mm}
\item What are the static features of design Hamiltonians such as thermal or quantum phases?
\end{enumerate}
The question I is the most interesting one, where we could use the methods developed in the random matrix theory~\cite{M1990}.
A natural candidate of design Hamiltonians satisfying all the three conditions of the conjecture may be $H_{\rm local GUE}=\sum_{\langle i, j \rangle} h_{ij}$ where each local term $h_{ij}$ is drawn randomly and independently from the so-called \emph{Gaussian unitary ensemble}~\cite{M1990} and the summation is taken over all neighbouring qubits.
We expect that $H_{\rm local GUE}$ generates a unitary design after some time although it may also be possible that it does not due to the many-body localization.
The question II is important to clarify the roles of design Hamiltonians in black hole information science and quantum chaos. As design Hamiltonians are based on unitary designs, it suffices to investigate explicit relations between unitary designs and scrambling or the OTO correlators. The relation between unitary designs and the OTO correlators is recently addressed and is clarified in Ref.~\cite{RB2016}.
The question III is not only of theoretical interest but also of practical importance because it determines the number of random bits necessary to construct design Hamiltonians.
To address this question, it is needed to relax the definition of design Hamiltonians to exclude the Poincar\'e recurrence time as we have mentioned in Section~\ref{Sec:DH}.
Note that, since the support of unitary $t$-designs on $N$ qubits should contain at least $O(2^{2tN})$ unitaries~\cite{RS2009}, the ensemble should contain at least the same number of Hamiltonians.  Finally, as design Hamiltonians are certain types of disordered Hamiltonians, it is natural to expect that they have special static properties, which is the question IV.
A static property of the above random Hamiltonian $H_{\rm local GUE}$ was numerically studied from the viewpoint of distributions in a state space, and evidences of phase transitions were obtained~\cite{NO2014}. 
However, as $H_{\rm local GUE}$ is not yet shown to be a design Hamiltonian and no time-independent design Hamiltonians have been found yet, it would be more realistic to start with investigating static properties of the Hamiltonian $H_Z$ of $H_{XZ}$, which has similarity to many-body localised systems, and their dependence on $t$.

\section{Acknowledgements}
The authors are grateful to S. Di Martino, C. Morgan and T. Sasaki 
for helpful discussions.
The authors also thank B. Yoshida for fruitful discussions and for telling us recent progresses on scrambling and quantum chaos, and C. Gogolin for pointing out the possibility of using our construction for quantum metrology. 
This work is supported by CREST, JST, Grant No. JPMJCR1671.
YN is a JSPS Research Fellow and is supported by JSPS KAKENHI Grant Number 272650.
AW and CH are supported by the Spanish MINECO, Projects No. FIS2013-40627-P and No. FIS2016-80681-P,
and CH by FPI Grant No. BES-2014-068888, as well as by the Generalitat de Catalunya, CIRIT project no. 2014 SGR 966.
AW is further supported by the European Commission (STREP ``RAQUEL''), the  European Research Council (Advanced Grant ``IRQUAT'').

\bibliographystyle{unsrt}

\bibliography{Bib}

\appendix

\section{Proof of Lemma~\ref{Lemma:additionalU}} \label{Ap:additionalU}
Here, a simple proof of Lemma~\ref{Lemma:additionalU} is given.

\begin{Proof}
As an $\epsilon$-approximate unitary $t$-design $U$ satisfies $|\!|  \mathcal{G}^{(t)}_{U\sim \nu} - \mathcal{G}^{(t)}_{U \sim {\sf H}}  |\! |_{\diamond} \leq \epsilon$, we have
\begin{align}
|\!|  \mc{G}^{(t)}_{V} \circ \mathcal{G}^{(t)}_{U\sim \nu} -  \mc{G}^{(t)}_{U \sim {\sf H}}  |\! |_{\diamond}
&=
|\!| \mc{G}^{(t)}_{V} \circ (  \mathcal{G}^{(t)}_{U\sim \nu} - \mc{G}^{(t)}_{U \sim {\sf H}} ) |\! |_{\diamond}\\
&\leq 
|\!| \mc{G}^{(t)}_{V} |\! |_{\diamond} |\!|  \mathcal{G}^{(t)}_{U\sim \nu} - \mc{G}^{(t)}_{U \sim {\sf H}} |\! |_{\diamond}\\
&\leq 
\epsilon ,
\end{align}
where we used the unitary invariance of the Haar measure in the first line, 
and a fact that $\mc{G}^{(t)}_{V}$ is a completely-positive and trace-preserving map in the last line. 
This implies that $VU$ is also an $\epsilon$-approximate unitary $t$-design.
The proof for $UV$ is similarly obtained.$\hfill \blacksquare$
\end{Proof}

\section{Proof of Theorem~\ref{Thm:TPEtoDESIGN}} \label{Sec:ProofTPEtoD}
Here, we provide a proof of Theorem~\ref{Thm:TPEtoDESIGN}, which follows almost directly from the following simple lemma.

\begin{Lemma} \label{Lemma:trivial}
{\it For any unitary $U$, it holds that $U^{\otimes t,t} = P_0 + (I - P_0)U^{\otimes t,t} (I - P_0)$.}
\end{Lemma}

\begin{Proof}
Using $\ket{\Psi_{\pi}} = I \otimes V(\pi) \ket{\Phi}$, we have for any $\pi \in S_t$ that 
\begin{align}
U^{\otimes t,t} \ket{\Psi_{\pi}} &= U^{\otimes t} \otimes U^{* \otimes t}V(\pi) \ket{\Phi}\\
&= U^{\otimes t}  \otimes V(\pi) U^{* \otimes t} \ket{\Phi}\\
&=   U^{\otimes t} U^{\dagger \otimes t}  \otimes V(\pi) \ket{\Phi}\\
&=  I  \otimes V(\pi) \ket{\Phi}\\
&= \ket{\Psi_{\pi}}, \label{Eq:2m;kl45}
\end{align}
where we have used the fact that $U^{* \otimes t}$ commutes with $V(\pi)$ in the second line and the property of the maximally entangled state in the third line.
This implies that $(I- P_0) U^{\otimes t,t} P_0 =0$. Replacing $U$ with $U^{\dagger}$ in Eq.~\eqref{Eq:2m;kl45}, we also have $(I- P_0) U^{\dagger \otimes t,t} P_0 =0$, implying $P_0 U^{\otimes t,t} (I-P_0)= 0$. Hence, we obtain
$U^{\otimes t,t}=P_0 + (I-P_0) U^{\otimes t,t}(I-P_0)$.  $\hfill \blacksquare$
\end{Proof}

\begin{Proof}[Theorem~\ref{Thm:TPEtoDESIGN}]
To prove Theorem~\ref{Thm:TPEtoDESIGN}, let $\nu$ be a quantum $(\eta,t)$-TPE, satisfying
\begin{equation}
|\! | \mathbb{E}_{U \sim \nu}[U^{\otimes t,t}] - \mathbb{E}_{U \sim {\sf H}}[U^{\otimes t,t}] |\! |_{\infty} \leq \eta.
\end{equation}
Applying Lemma~\ref{Lemma:trivial} to all the unitaries in $\mathbb{E}_{U \sim \nu}[U^{\otimes t,t}]$, we have 
\begin{equation}
\mathbb{E}_{U \sim \nu}[U^{\otimes t,t}] = P_0 + (I-P_0)\mathbb{E}_{U \sim \nu}[U^{\otimes t,t}](I-P_0). \label{Eq:2;,3}
\end{equation}
Due to Lemma~\ref{Lemma:A}, which reads $P_0=\mathbb{E}_{U \sim {\sf H}}[U^{\otimes t,t}]$, the quantum TPE $\nu$ satisfies that
\begin{equation}
|\! | (I-P_0) \mathbb{E}_{U \sim \nu}[U^{\otimes t,t}] (I-P_0)|\! |_{\infty} \leq \eta. \label{Eq:23no;}
\end{equation}
Let $\nu^{\ell}$ be a measure corresponding to that of the $\ell$ iterations of the quantum TPE $\nu$. Then,
\begin{align}
|\! |\mc{G}^{(t)}_{U \sim \nu^{\ell}} - \mc{G}^{(t)}_{U \sim {\sf H}} |\! |_{\diamond} &\leq d^t |\! |\mc{G}^{(t)}_{U \sim \nu^{\ell}} - \mc{G}^{(t)}_{U \sim {\sf H}} |\! |_{2 \rightarrow 2}\\
&=d^t|\! | \mathbb{E}_{U \sim \nu^{\ell}}[U^{\otimes t,t}] - \mathbb{E}_{U \sim {\sf H}}[U^{\otimes t,t}]|\! |_{\infty} \\ 
&=
d^{t} |\! | \bigl(\mathbb{E}_{U \sim \nu}[U^{\otimes t,t}] \bigr)^{\ell} - \mathbb{E}_{U \sim {\sf H}}[U^{\otimes t,t}]|\! |_{\infty} \\
&=
d^{t} |\! |\bigl((I-P_0) \mathbb{E}_{U \sim \nu}[U^{\otimes t,t}] (I-P_0) \bigr)^{\ell}|\! |_{\infty}\\
&\leq
d^{t} |\! |(I-P_0) \mathbb{E}_{U \sim \nu}[U^{\otimes t,t}] (I-P_0) |\! |_{\infty}^{\ell}\\
&\leq
d^{t} \eta^{\ell}.
\end{align}
Here, the first line is due to the inequality that $|\! |\mc{E} |\! |_{\diamond} \leq D |\! |\mc{E} |\! |_{2 \rightarrow 2}$ for any superoperators $\mc{E}$ acting on a $D$-dimensional system, the third line is obtained due to the independence of the measure of each iteration, the fourth line is from Eq.~\eqref{Eq:2;,3}, and the last line is from Eq.~\eqref{Eq:23no;}.
This implies that $\ell$ iterations of a quantum $(\eta,t)$-TPE is an $\epsilon$-approximate unitary $t$-design if $d^{t} \eta^{\ell} \leq \epsilon$. $\hfill \blacksquare$
\end{Proof}

\section{Proof of Lemma~\ref{Lemma:FT} } \label{Sec:FourierType}

Here, we prove Lemma~\ref{Lemma:FT} about the Fourier-type bases.

\begin{Proof}[Lemma~\ref{Lemma:FT}]
When a pair of two bases is that of arbitrary basis and its Fourier basis, it is clear that $\theta_{k \alpha} = \frac{2 \pi k \alpha}{d}$ and the additive operation in the index is given by an addition modulo $d$. It can be easily checked that $[0,d-1]$ is an additive group with respect to the modular addition.

When the pair is given by the Pauli-$X$ and -$Z$ bases, using the binary representation such as $\alpha=\alpha_1 \cdots \alpha_N$ $(\forall j \in[1,N], \alpha_j \in \{0,1\})$,
the Pauli-$X$ and -$Z$ bases can be represented by 
\begin{equation}
\ket{\alpha}_X = \bigotimes_{j=1}^N \ket{\alpha_j}_X,\ \ \ \ \ \ 
\ket{k}_Z= \bigotimes_{j=1}^N \ket{k_j}_Z,
\end{equation}
respectively. Using a fact that $_Z\!\braket{k_j}{\alpha_j}_X = {}_X\!\braket{\alpha_j}{k_j}_Z$ is equal t to $1/\sqrt{2}$ if $(\alpha_j, k_j) =(0,0), (0,1), (1,0)$ and is equal to $-1/\sqrt{2}$ if $(\alpha_j,k_j) =(1,1)$, we have $\theta_{k \alpha}=\pi \sum_{j=1}^N \delta_{k_j 1} \delta_{\alpha_j 1}$, leading to
\begin{align}
\exp \bigl[i (\theta_{k \alpha} + \theta_{l \alpha}) \bigr] &= \exp\bigl[i \pi \sum_{j=1}^N (\delta_{k_j 1} + \delta_{l_j 1})\delta_{\alpha_j 1}\bigr]\\
&=\exp \bigl[i \pi \sum_{j=1}^N \delta_{k_j + l_j, 1}\delta_{\alpha_j 1} \bigr]\\
&=\exp \bigl[i \theta_{k \oplus l, \alpha} \bigr],
\end{align}
where $\oplus$ is a bitwise XOR, defined by $a \oplus b = 0$ when $a=b$ and otherwise $1$ for binary numbers $a$ and $b$, 
and is the additive operation in the index, making $[0,d-1]$ an additive group.
$\hfill \blacksquare$
\end{Proof}

\section{Proof of Lemma~\ref{Lemma:LP}} \label{Sec:PLP}

Here, we prove Lemma~\ref{Lemma:LP} which connects the achievability of quantum-TPE with random diagonal circuits and the local permutation check problem.

\begin{Proof}[Lemma~\ref{Lemma:LP}]
Let ${\sf RDC}(\mc{I})$ be the probability measure of RDC$(\mc{I})$. We denote the averaged operators $\mbb{E}_{D ^Z\sim {\sf RDC}(\mc{I})}[(D^Z)^{\otimes t,t}]$ and $\mbb{E}_{D^Z \sim {\sf D}_Z}[(D^Z)^{\otimes t,t}]$ by $Q_Z$ and $P_Z$, respectively. 
There exists a projector $R_Z$ diagonal in the Pauli-$Z$ basis such that $Q_Z=P_Z + R_Z$ because $Q_Z P_Z = P_Z Q_Z = P_Z$ and $Q_Z$ is a projector diagonal in the Pauli-$Z$ basis. 
Denoting $H_N^{\otimes t,t} Q_Z H_N^{\otimes t,t}$ by $Q_X$, where $H_N:=H^{\otimes N}$ is the Hadamard transformation on $N$ qubits, and similarly decomposing it into $P_X + R_X$ ($P_X:=H_N^{\otimes t,t} P_Z H_N^{\otimes t,t}$ and $R_X:=H_N^{\otimes t,t} R_Z H_N^{\otimes t,t}$), we have
\begin{align}
|\!| Q_Z Q_X Q_Z - P_0  |\!|_{\infty}
&=|\!| P_Z P_X P_Z - P_0 + R_Z P_X P_Z+ Q_Z P_X R_Z + Q_Z R_X P_Z + Q_Z R_X R_Z  |\!|_{\infty}\\
&\leq |\!| P_ZP_XP_Z - P_0  |\!|_{\infty} + 2|\!| P_X R_Z |\!|_{\infty} +|\!| R_X P_Z |\!|_{\infty}
+|\!| R_X R_Z |\!|_{\infty}\\
&\leq \eta + 2|\!| P_X R_Z |\!|_{\infty} +|\!| R_X P_Z |\!|_{\infty}
+|\!| R_X R_Z |\!|_{\infty}, \label{Eq:ZZ}
\end{align}
where we used Theorem~\ref{Thm:main} in the last line.

We denote by $\mc{W}_{Z}$ a set of $(\mbf{k}_1, \mbf{k}_2) \in \mc{N} \times \mc{N}$ such that $\bra{\mbf{k}_1,\mbf{k}_2} R_Z \ket{\mbf{k}_1,\mbf{k}_2}=1$. 
Using an upper bound of the operator norm by the row norm and using the fact that $| \bra{\mbf{l}_1,\mbf{l}_2} P_X \ket{\mbf{k}_1, \mbf{k}_2} | = (\tr P_X)/2^{2tN} \leq t!/2^{tN}$ for any $(\mbf{k}_1,\mbf{k}_2)$ and $(\mbf{l}_1,\mbf{l}_2)$, we obtain
\begin{align}
|\!| R_X P_Z |\!|_{\infty}=|\!| P_X R_Z |\!|_{\infty} \leq \max_{(\mbf{l}_1,\mbf{l}_2) \in \mc{W}_{Z}}
\sum_{(\mbf{k}_1,\mbf{k}_2) \in \mc{W}_{Z}}
\biggl|  \bra{\mbf{l}_1,\mbf{l}_2} P_X \ket{\mbf{k}_1,\mbf{k}_2}  \biggr| \leq 
\frac{t!}{2^{tN}} | \mc{W}_Z|. \label{Eq:AA}
\end{align}
Similarly, we have
\begin{align}
|\!| R_X R_Z |\!|_{\infty} 
&\leq \max_{(\mbf{l}_1,\mbf{l}_2) \in \mc{W}_{Z}}
\sum_{(\mbf{k}_1,\mbf{k}_2) \in \mc{W}_Z}
\biggl|  \bra{\mbf{l}_1,\mbf{l}_2} R_X \ket{\mbf{k}_1,\mbf{k}_2}  \biggr|
\leq
\biggl(\frac{| \mc{W}_Z| }{2^{tN}} \biggr)^2. \label{Eq:BB}
\end{align}
Substituting Eqs.~\eqref{Eq:AA} and ~\eqref{Eq:BB} into Eq.~\eqref{Eq:ZZ}, we obtain
\begin{align}
|\!|Q_Z  Q_X Q_Z - P_0  |\!|_{\infty}
&\leq \eta +3 t! \frac{| \mc{W}_Z|}{2^{tN}}  +\biggl(\frac{| \mc{W}_Z| }{2^{tN}} \biggr)^2.
\end{align}

We finally show that $| \mc{W}_Z| =\Lambda(\mc{I})$.
Note that $\Lambda(\mc{I})$ is the number of $(K, K') \in \{0,1 \}^{t N} \times \{0,1 \}^{t N}$ such that $K$ is not a row permutation but is an $\mc{I}$-local permutation of $K'$.
We first express each $k_s \in \mbf{k}$ in binary such as $k_s = k_{s1} \cdots k_{sN}$ and define a $t \times N$ matrix $K$ with elements in $\{0,1\}$ corresponding to $\mbf{k}$, 
\begin{equation}
K := \begin{pmatrix}
k_{11} & k_{12} & \cdots & k_{1N} \\ 
\vdots & \vdots & \ddots & \vdots\\
k_{t1} & k_{t2} & \cdots & k_{tN} \\
\end{pmatrix},
\end{equation}
where $k_{sm} \in \{0,1\}$. 
Using this notation and noting that the $Z$-basis is real, the state $\ket{ \mbf{k}, \mbf{k}'^*}$ is expressed as $\ket{ K, K'}$.
A random diagonal gate in RDC$(\mc{I})$ applied on qubits in $I \in \mc{I}$ corresponds to,
after taking the tensor product and the average, an projector onto ${\rm span}\{ \ket{K, K'}: \Omega(K_I) =\Omega(K'_I)\}$, where $\Omega$ is a canonical map that rearranges $|I|$-bit sequences $\{ K_{s,I} \}_{s \in [1,t]}$ in ascending order.
Thus, we have 
\begin{equation}
\bra{K, K'}Q_Z \ket{K, K'} =
\begin{cases}
1 & \text{if $\forall I \in \mc{I}$, $\Omega(K_{I}) = \Omega(K'_{I})$}, \\
0 & \text{otherwise}.
\end{cases}
\end{equation}
Note that the off-diagonal elements of $Q_Z$ are always zero because it is diagonal in the $Z$ basis.
We also have 
\begin{equation}
\bra{K, K'}P_Z \ket{K, K'}
=\begin{cases}
1 & \text{if $K$ is a row permutation of $K'$}, \\
0 & \text{otherwise}.
\end{cases}
\end{equation}
From these two equations, it is clear that $R_Z = Q_Z -P_Z$ satisfies that
$\bra{K, K'}R_Z \ket{K, K'}=1$ if and only if 
$K$ is not a row permutation but is an $\mc{I}$-local permutation of $K'$. Otherwise $\bra{K, K'}R_Z \ket{K, K'}=0$. This implies $| \mc{W}_Z| = \Lambda(\mc{I})$.
 $\hfill \blacksquare$
\end{Proof}

\section{Proof of Lemma~\ref{Lemma:counting}} \label{Sec:ProofCounting}

Here, we prove Lemma~\ref{Lemma:counting}.

\begin{Proof}[Lemma~\ref{Lemma:counting}]
Let $i \in [1,t]$ and $\vec{e}_i$ be a vector with elements in $\{0,1\}$ where only the $i$th element is $1$:
\begin{equation}
\vec{e}_i = (0,\dots,0,\place{i}{1},0,\dots,0)^T.
\end{equation}
Then, for any $i$, there exists a vector $\vec{v}_i \in \{ 0,1 \}^t$ such that both $\vec{v}_i$ and $\vec{v}_i + \vec{e}_i$ are contained in $S$.
This is for the following reason: if there is no such pair of $\vec{v}_i$ and $\vec{v}_i + \vec{e}_i$, it implies that a pair of vectors, which have different values only at the $i$th element, is not contained in $S$. This results in $|S| \leq 2^{t-1}$, which is in contradiction to the assumption that $|S| > 2^{t-1}$.

As $\vec{v}_i + \vec{e}_i \in S \subset \{0,1\}^t$, the $i$th element of $\vec{v}_i$ is $0$.
Hence, we have $\vec{v}_i \cdot \vec{e}_i=0$, implying that
\begin{equation}
O\vec{e}_i  \cdot O(\vec{v}_i + \vec{e}_i) = \vec{e}_i  \cdot \vec{v}_i + \vec{e}_i  \cdot \vec{e}_i
= \vec{e}_i  \cdot\vec{e}_i = 1.
\end{equation}
It is also trivial that $O\vec{e}_i \cdot O\vec{e}_i = 1$ and that $O\vec{e}_i \in \{-1,0,1\}^t$, which follows from an identity $O\vec{e}_i = O(\vec{v}_i + \vec{e}_i) - O \vec{v}_i$ and a fact that both $O(\vec{v}_i + \vec{e}_i)$ and $O \vec{v}_i$ are in $\{0,1\}^t$.
From these three relations and again $O(\vec{v}_i + \vec{e}_i) \in \{0,1\}^t$, we conclude
\begin{equation}
O\vec{e}_i = \vec{e}_j = (0,\dots,0,\place{j}{1},0,\dots,0)^T
\end{equation}
for some $j \in [1,t]$.
Because $O$ is invertible, this implies that $O$ is a permutation matrix. 
%Additionally, we can conclude that $S$ is the entire hypercube.
$\hfill \blacksquare$
\end{Proof}

\end{document}